\documentclass[english,draftcls,onecolumn]{IEEEtran}
\usepackage[T1]{fontenc}
\usepackage[latin9]{inputenc}
\usepackage{array}
\usepackage{amsmath}
\usepackage{amsthm}
\usepackage{amssymb}
\usepackage{graphicx}

\makeatletter

\providecommand{\tabularnewline}{\\}

  \theoremstyle{plain}
  \newtheorem{thm}{\protect\theoremname}
  \theoremstyle{plain}
  \newtheorem{lem}{\protect\lemmaname}
  \theoremstyle{remark}
  \newtheorem{rem}{\protect\remarkname}

\makeatother

\usepackage{babel}
\providecommand{\lemmaname}{Lemma}
\providecommand{\remarkname}{Remark}
\providecommand{\theoremname}{Theorem}

\begin{document}

\title{Degrees of Freedom of the MIMO $2 \times 2$ Interference Network
with General Message Sets}

\author{Yao Wang and Mahesh K. Varanasi,\textit{ Fellow, IEEE} \thanks{This work was supported in part by NSF Grant 1423657. The material was presented in part at the IEEE Intnl. Symp. Information Theory (ISIT) 2015, Hong Kong \cite{Wang2015}. The authors are with the Electrical, Computer and Energy Engineering Department of the University of Colorado, Boulder, CO 80309-0425.}}
\maketitle
\begin{abstract}
We establish the degrees of freedom (DoF) region for the multiple-input
multiple-output (MIMO) two-transmitter, two-receiver ($2 \times 2$)
interference network with a general message set, consisting of nine messages, 
one for each pair of a subset of  transmitters at which that message is known and a subset of receivers where that message is desired. An outer bound on the general nine-message $2 \times 2$ interference network is obtained and then it is shown to be tight, establishing the DoF region for the most general antenna setting wherein all four nodes have an arbitrary number of antennas each. The DoF-optimal scheme is applicable to the  MIMO $2 \times 2$ interference network with constant channel coefficients, and hence, \emph{a fortiori}, to time/frequency varying channel scenarios. 


 In particular, a linear precoding scheme is proposed that can achieve all the DoF tuples in the DoF region. In it, the precise roles played by transmit zero-forcing, interference alignment, random beamforming, symbol extensions and asymmetric complex signaling (ACS) are delineated. For instance, we identify a class of antenna settings in which ACS is required to achieve the fractional-valued corner points. 
 
Evidently, the  DoF regions of all previously unknown cases of the $2 \times 2$ interference network with a subset of the nine-messages are established as special cases of the general result of this paper. In particular, the DoF region of the well-known four-message (and even three-message) MIMO $X$ channel is established. This problem had remained  open despite previous studies which had found inner and outer bounds that were not tight in general. Hence, the DoF regions of all special cases obtained from the general DoF region of the nine-message 2$\times$2 interference network of this work that include at least three of the four $X$ channel messages are new, among many others. Our work sheds light on how the same physical $2 \times 2$ interference network could be used by a suitable choice of message sets to take most advantage of the channel resource in a flexible and efficient manner.\end{abstract}


\section{Introduction}

In order to design communication systems that can flexibly and efficiently
handle the complex signaling requirements of modern applications,
such as in the delivery phase of caching systems over wireless interference
channels, it may be necessary to offer multiple physical layer modes
that allow for the transmission of some or all of multiple unicast,
multiple multicast, multiple broadcast (i.e., $X$-channel), and/or
cooperative/cognitive/common messages. In this paper, rather than
considering each such transmission mode in isolation, we study the
unified setting in which any subset (including all) such messages
can be transmitted simultaneously over the MIMO $2 \times 2$ interference network. For this simple network,
depending on the subset of the two transmitters at which a message
is known, and the subset of the two receivers where it is desired,
there are nine possible messages in the general message set. For this
fully general message set, the associated nine-dimensional DoF region
of the MIMO $2 \times 2$ interference network is established herein.

The most studied and also the best understood setting of the $2 \times 2$
interference network is the two-unicast setting, referred to in the literature as the 
interference channel \cite{ElGamal2011}, in which each transmitter has a private message for its single distinct intended receiver (cf. \cite{Jafar2007,Karmakar2012,Karmakar2013} and the references therein). 
In particular, the DoF region
of the two-user MIMO interference channel was found in \cite{Jafar2007} and more refined characterizations in terms of generalized degrees of freedom and constant bit-gap to capacity were found in \cite{Karmakar2012} and \cite{Karmakar2013}, respectively.

The four private message case, which can be thought of
as a two-broadcast network, more commonly known as the $X$ channel, 
allows for the transmission of a private message to each of the two
receivers from each transmitter. The now well-known, and more broadly applicable, 
linear precoding technique known as {\em interference alignment} is needed to achieve the DoF in some cases. With its use, the MIMO $X$ channel was shown in \cite{MMK2008IA,jafar2008degrees}
to achieve higher sum DoF than the MIMO interference channel. For example, when
all transmitters and receivers are equipped with the same number,
$M$, of antennas, the two-user MIMO interference channel has a sum
DoF of $M$, while the MIMO $X$ channel can achieve a sum DoF of
$\frac{4}{3}M$ for $M>1$, achievable with interference alignment. The key idea
(when $M $ is a multiple of 3) is that by aligning undesired signals
(i.e., interference) from the two transmitters into the same subspace
at a receiver, one can maximize the desired signal dimensions at that
receiver. In \cite{jafar2008degrees}, an outer bound on the DoF region
of the MIMO $X$ channel is given based on the sum rate outer bound
of the embedded MAC, BC and $Z$ channels in the $X$ channel. Moreover,
\cite{jafar2008degrees} gives an achievability scheme based on
interference alignment and presents an achievable DoF region that
is given as the convex hull of all integer-valued degrees of freedom
within that outer bound region. But these inner and the outer bounds
of \cite{jafar2008degrees}
are not identical. However, using interference alignment over multi-letter
extensions of the MIMO $X$ channel, it was shown that the outer bound
is tight (including non-integer corner points) when all nodes have
equal number of antennas $M$, when $M>1$. In the context of the
general MIMO $X$ channel with an arbitrary numbers of antennas at the
four terminals, \cite{jafar2008degrees} claims that the DoF outer
bound region obtained therein is tight in ``most cases'', but a
precise statement and proof of this claim is not provided. Later,
the authors of \cite{cadambe2010interference} introduced a novel
technique named asymmetric complex signaling (ACS). By allowing the
inputs to be complex but not circularly symmetric and using an alternative
representation of the channel models in terms of only real quantities,
the problem is transformed to delivering real messages over channels with real-valued coefficients.
Consequently, it was shown that the 2-user single-input, single-output (SISO) $X$ channel with constant channel
coefficients achieves the outer bound of $\frac{4}{3}$ DoF. However, it remained an open problem as to whether the outer bound of \cite{jafar2008degrees} is tight for any of the multiple antenna
cases. For instance, the problem remained open as to whether there are other scenarios in which ACS is required in addition to multi-letter extensions, as did the problem of identifying cases in which just multi-letter extensions suffice to achieve the outer bound. More recently, it was shown in \cite{Agustin2012} that the outer bound on the {\em sum}
DoF for the MIMO $X$ channel (with generic channel coefficients)
derived in \cite{jafar2008degrees} is tight for any antenna configuration.
The work in \cite{Agustin2012} proposes a linear precoding method
based on the generalized singular value decomposition (GSVD), and
with the aid of computational experiments, the authors of \cite{Agustin2012}
offer a conjecture that the outer bound {\em region} obtained in
\cite{jafar2008degrees} is also tight. The general DoF region result
of this paper for the MIMO $2 \times 2$ interference network with
nine distinct messages, when specialized to the four private-message
MIMO $X$ channel, settles this conjecture in the affirmative. It
therefore also expands on, and makes precise, the claim in \cite{jafar2008degrees}. 
The outer bound on the DoF region of \cite{jafar2008degrees} is indeed tight.

Besides the aforementioned MIMO interference and $X$ channels (and
its embedded MAC and/or BC), in which only private messages are considered
(see also \cite{Karmakar2012,Karmakar2013,Niesen-Maddah2013}), the
$2 \times 2$ interference network can work in various other modes
if common messages, multicast messages and/or transmitter cognition
are allowed. For example, if both transmitters share the same three
messages, and each receiver demands one of the first two messages
while both demand the third, we have what is known as the broadcast channel
with private and common messages (BC-CM) \cite{Ekrem-Ulukus2012,Geng-Nair-bccm-2014}.
On the other hand, if each transmitter has a private message, and
both receivers demand both of the messages, the system works as a
compound multiple access channel (C-MAC) \cite{Ahlswede74}. If there
are two private messages as in the interference channel and there
is a common message known by both transmitters and also demanded by
both receivers, the network is known as the interference channel with
common message (IC-CM) \cite{Jiang2008,RomeroMV13}. The network is
referred to as a cognitive $X$ channel in \cite{jafar2008degrees} if
there are four independent messages to be sent as in the $X$ channel, but with
one of the four messages known at both transmitters. A new three-message
setting could be defined in which one transmitter has 2 messages,
each intended for a distinct receiver, and a third shared message
that is known to both transmitters and desired at one of the receivers.
Interpreting the second transmitter as a relay, such a setting could
be described as a broadcast channel with a partially cognitive relay
(BC-PCR). A six-message cognitive $X$ channel could be defined as
having the four private messages as in the $X$ channel as well as
two more messages that are known to both transmitters with each desired
at a distinct receiver. Evidently, based on different message sets,
the $2 \times 2$ interference network can represent many different
settings and potential applications. 
\begin{figure*}[t]
\centering{}\includegraphics[height=3cm]{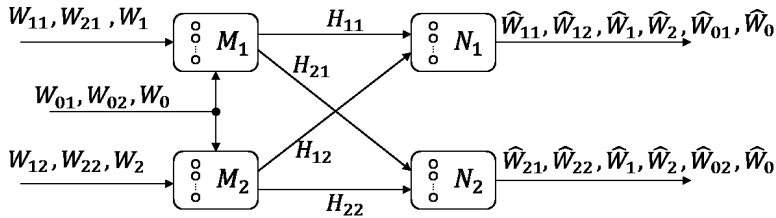}\caption{\label{fig:IC_GM_model-1} The $ 2 \times 2$ Interference Network with General Message Set}
\end{figure*}


\emph{Notation:} $\textrm{co(A)}$ is the convex hull of set $\textrm{A}$,
$\mathbb{R}_{+}^{n}$ and $\mathbb{Z}_{+}^{n}$ denote the set of
non-negative $n$-tuples of real numbers and integers, respectively. $(x)^{+}$
represents the larger of the two numbers, $x$ and $0$. $A\otimes B$ denotes
the Kronecker product of matrix $A$ and $B$. $[A\,\,B]$ means the
horizontal concatenation of matrix $A$ and $B$, and $[A;B]$ means
the vertical concatenation of matrix $A$ and $B$. $\textrm{Re}(A)$
and $\textrm{Im}(A)$ denote the real part and imaginary part of complex
matrix $A$, respectively. $\mathcal{N}(A)$ denotes the null space
of the linear transformation $A$. $\mathrm{Span}(V)$ denotes the
subspace spanned by the column vectors of matrix $V$.

\section{\label{sec:System-Model}System Model}

We consider the complex Gaussian network with two transmitters and
two receivers, as it is shown in Figure \ref{fig:IC_GM_model-1}.
The two transmitters are equipped with $M_{1}$, $M_{2}$ antennas
respectively, and the two receivers are equipped with $N_{1}$, $N_{2}$
antennas respectively. We denote the channel between transmitter $t$
and receiver $r$ as the $N_{r}\times M_{t}$ complex matrix $H_{rt}$
and assume all channels to be generic, i.e., all the channel coefficient
values are drawn independently from a continuous probability distribution.
The channel is assumed to be constant over the duration of communication
and all channel coefficients are perfectly known at all transmitters
and receivers. The received signal at receiver $r\,(r=1,2)$ is given
by $Y_{r}=H_{r1}X_{1}+H_{r2}X_{2}+Z_{r}$, where $X_{t}\,(t=1,2)$
is the $M_{t}\times1$ input vector at transmitter $t$, $Z_{r}$
is the $N_{r}\times1$ additive white Gaussian noise (AWGN) vector
at receiver r.

General message sets are considered in this paper. For $2 \times 2$
interference network, there are at most nine possible messages classified
by different sources and destinations. We index them as $W_{11}$,
$W_{12}$, $W_{21}$, $W_{22}$, $W_{01}$, $W_{02}$, $W_{0}$, $W_{1}$
and $W_{2}$, as shown in Figure \ref{fig:IC_GM_model-1}. $W_{rt}\,(r,t=1,2)$
is a private message sent from transmitter $t$ to receiver $r$;
$W_{0r}\,(r=1,2)$ is a common message transmitted cooperatively from
both transmitters to receiver $r$; $W_{t}\,(t=1,2)$ is a multicast
message transmitted from transmitter $t$ and demanded by both receivers
simultaneously; $W_{0}$ is a common multicast message transmitted
cooperatively from both transmitters and demanded by both receivers.

Assume the total power across all transmitters to be equal to $\rho$
and indicate the message set size by $\left|W(\rho)\right|$. For
codewords occupying $t_{0}$ channel uses, the rates $R(\rho)=\frac{\log\left|W(\rho)\right|}{t_{0}}$
are achievable if the probability of error for all nine messages can
simultaneously be made arbitrarily small by choosing appropriately
large $t_{0}$. The capacity region $\boldsymbol{C}(\rho)$ of the
MIMO $2 \times 2$ interference network with general message sets
is the set of all achievable rate-tuples $\boldsymbol{R}(\rho)$=$\left(R_{11}\left(\rho\right),R_{12}\left(\rho\right),...,R_{0}\left(\rho\right)\right)$.
Define the degrees of freedom region $\mathbb{D}$ for MIMO  $2 \times 2$
interference network with general message sets as
\[
\begin{array}{c}
\mathbb{D}\triangleq\Biggl\{(d_{11},d_{12},...,d_{0})\in\mathbb{R}_{+}^{\mathsf{E}}:\,\forall(\omega_{11},\omega_{12},...,\omega_{0})\in\mathbb{R}_{+}^{\mathsf{E}}\\
\underset{x\in\mathsf{E}}{\sum}\omega_{x}d_{x}\leq\underset{\rho\rightarrow\infty}{\limsup}\left[\underset{\boldsymbol{R}(\rho)\in\boldsymbol{C}(\rho)}{\sup}\frac{\underset{x\in\mathsf{E}}{\sum}\omega_{x}R_{x}(\rho)}{\log(\rho)}\right]\Biggr\}.
\end{array}
\]
\\
\\
where $\mathsf{E}=\left\{ 11,12,21,22,01,02,1,2,0\right\} $.

This definition is the general message set counterpart of the one
provided in \cite{jafar2008degrees} for the MIMO X channel. Note that 
$\mathbb{D}$ is a closed convex set.

In the following section, we consider first the previously studied MIMO $X$ channel, for which the best inner and outer bounds of \cite{jafar2008degrees,cadambe2010interference,Agustin2012} known to date are not coincident in general. The MIMO $X$ channel provides the context in which to introduce the notation used in this paper and all the relevant linear precoding techniques, namely, 
zero-forcing, interference alignment, symbol extension, and ACS. We provide a class of antenna configurations for which, among linear schemes, ACS is required  and is sufficient, along with multi-letter extensions and the other linear precoding techniques, to achieve all fractional DoF corner points for those antenna configurations. More generally, we show that the use of linear precoding techniques including symbol extensions and ACS, whether ACS is required or not, are sufficient to achieve any corner point of the DoF region regardless of the antenna configuration. The DoF region of the general nine-message problem is established in \ref{sec:Main-Result}.


\section{\label{sec:The-MIMO-X}The MIMO $X$ Channel}

The  MIMO $X$ channel is an important special case of the $2\times 2$ interference network in which only the four private messages, namely, $W_{11}$,
$W_{12}$, $W_{21}$, $W_{22}$,  are present. Hence the message index set in this case is $\sf{E} = \{11, 12, 21, 22\}$. Each of these four messages is
intended for one of the two receivers and is a source of interference
to the other receiver. 

We start by stating the DoF region of the MIMO $X$ channel.
\begin{thm}
The DoF region of the MIMO $X$ channel with constant generic channel coefficients is (with probability one)
\begin{align*}
\mathbb{D}_{X}=\lbrace & (d_{11},d_{21},d_{12},d_{22})\in\mathbb{R}_{+}^{4}:\\
 & d_{11}+d_{12}+d_{21}\leq\max(M_{1},N_{1}),\\
 & d_{11}+d_{12}+d_{22}\leq\max(M_{2},N_{1}),\\
 & d_{21}+d_{22}+d_{11}\leq\max(M_{1},N_{2}),\\
 & d_{21}+d_{22}+d_{12}\leq\max(M_{2},N_{2}),\\
 & d_{11}+d_{12}\leq N_{1},\,\,d_{21}+d_{22}\leq N_{2},\\
 & d_{11}+d_{21}\leq M_{1},\,\,d_{12}+d_{22}\leq M_{2}\rbrace.
\end{align*}
\label{thm-mimo-x}
\end{thm}
That the above DoF region is an outer bound for the DoF region of the MIMO $X$ channel is proved in Theorem 2 of \cite{jafar2008degrees}. The outer bounding inequalities result, respectively, from the embedded multiple-access
channel, broadcast channel and  $Z$ channels, in the MIMO
$X$ channel. The readers can refer to \cite{jafar2008degrees} for details.
Moreover, these outer bounds are generalized to the general nine-message problem
in Section \ref{sec:Outerbound-on-the}. 

The authors of \cite{jafar2008degrees} also provide a constructive
achievability proof to show that the convex hull of all the integer-valued
DoF-tuples in $\mathbb{D}_{X}$ is achievable. The techniques used
in the achievable scheme are zero-forcing, interference alignment
and random beamforming. Since these techniques are among the techniques used in our $2 \times 2$ interference network with general message sets problem, we provide a succinct account of them in Section \ref{sub:Zero-forcing-and-interference}, describing in the process, the notation used in this paper as well. The techniques of symbol extension and ACS are described in Sections \ref{sub-section:se-acs} and \ref{sub-section-acs} to follow.


\subsection{\label{sub:Zero-forcing-and-interference}Zero-forcing and interference
alignment}

Consider message $W_{11}$ as an example. If $M_{1}>N_{2}$, the null space
of channel $H_{21}$ is not empty. By transmitting some symbols of
message $W_{11}$ using the beamformers chosen from the null space
$\mathcal{N}(H_{21})$, we can zero-force these symbols at receiver
$R_{2}$ and thus introduce no interference to it. The maximum number
of such symbols that can be zero-forced is $(M_{1}-N_{2})^{+}$, which
is equal to the rank of $\mathcal{N}(H_{21})$. Similarly, we can
transmit, at most, $(M_{2}-N_{2})^{+}$ symbols of message $W_{12}$
via the nullspace of channel $H_{22}$ and zero-force them all at
their unintended receiver, $R_{1}$. Note that the
null space $\mathcal{N}(H_{21})$ and $\mathcal{N}(H_{22})$ are both
subspaces of the null space of the concatenated channel $[H_{21}\ H_{22}]$.
The remaining dimension of $\mathcal{N}([H_{21}\ H_{22}])$ is equal
to $A=(M_{1}+M_{2}-N_{2})^{+}-(M_{1}-N_{2})^{+}-(M_{2}-N_{2})^{+}$.
By choosing beamformers for message $W_{11}$ and $W_{12}$ jointly
from the rest of the subspace of $\mathcal{N}([H_{21}\ H_{22}])$, it is
possible to align this part of message $W_{11}$ and $W_{12}$ 
into the same subspace and thus reserve more dimensions for
the desired messages at receiver $R_{2}$, and the maximum number
of such pairs of streams is equal to $A$. If there are more
symbols of message $W_{11}$ left, they can be transmitted using random 
beamforming, which would create unavoidable interference at its unintended receiver.

Since the technique of zero-forcing is the more efficient in terms of
reducing interference than interference alignment, it is given the
highest priority when constructing precoding beamformers. Following that,
interference alignment is used to the extent possible, and following which
all of the remaining symbols are sent using random beamforming. The
beamformers for each private message is hence divided into three linearly
independent parts based on the precoding technique used. Here we use
superscript '\textit{Z}' to indicate a message is zero-forced at its
unintended receiver, '\textit{A}' to indicate a message
is aligned with another interference at their commonly unintended
receiver, and '\textit{R}' to indicate the remainder
of a certain message that is transmitted using random beamforming.
Hence a message $W_x $ for $ x \in \sf{E}$ (recall $ \sf{E} = \{11, 12, 21, 22\}$ for the MIMO $X$ channel) is split in general into three components or sub-messages, denoted $W_x^Z, W_x^A$ and $W_x^R$, with the number of symbols (dimensions) in each denoted as $d_x^Z$, $ d_x^A$ and $d_x^R$, respectively. In general, we use the notation $W_x^y$ and $d_x^y$ with $x \in \sf{E}$ and $y \in \{\textit{Z},\textit{A},\textit{R}\}$ for the component messages and dimensions, respectively. Similarly, the precoding matrix for any sub-message $W_x^y$ 
is denoted as $V_{x}^{y}$. Thus we have that $d_{ij}=d_{ij}^{Z}+d_{ij}^{A}+d_{ij}^{R}$ and let $V_{ij}$ denote the horizontally concatenated matrix $V_{ij}=[V_{ij}^{Z}\ V_{ij}^{A}\ V_{ij}^{R}]$, where $i,j=1,2$. It was shown in \cite{jafar2008degrees}
that any integer-valued DoF-tuple within the outer
bound can be divided into three such parts within the
decoding ability of the channels. It is thus achievable.

\subsection{Symbol extensions}
\label{sub-section:se-acs}

When a corner point of $\mathbb{D}_{X}$ is not integer-valued, it is rational-valued.
It is therefore natural to consider a multi-letter extension of the channels to obtain a larger but equivalent system with the corresponding corner point of the DoF region being integer-valued. The length of symbol extensions can be chosen to be the least common multiple of the denominators of all the fractional values. To this time-extended channel, the techniques of zero-forcing, alignment and random beamforming can be applied as described in the previous section. This was proposed in \cite{jafar2008degrees}.

Consider $T$ symbol extensions of the $X$ channel with complex and constant
(across time) channel coefficients. We have the equivalent $\widetilde{N}_{i}\times\widetilde{M}_{j}$
channel matrix $\widetilde{H}_{ij}$, in which $i,j=1,2$, $\widetilde{M}_{i}=T\cdot M_{i}$,
$\widetilde{N}_{i}=T\cdot N_{i}$, and
\begin{eqnarray}
\widetilde{H}_{ij} & = & I_{T}\otimes H_{ij}\nonumber \\
 & = & \left[\begin{array}{cccc}
H_{ij} & 0 & \ldots & 0\\
0 & H_{ij} & \ldots & 0\\
\vdots & \vdots & \ddots & \vdots\\
0 & 0 & \ldots & H_{ij}
\end{array}\right].\label{eq:Hij_time_ext-1}
\end{eqnarray}
Hence, we effectively have an $X$ channel with $\widetilde{M}_{j}$
antennas at the $j$th transmitter and $\widetilde{N}_{i}$ antennas
at the $i$th receiver and channel matrices $\widetilde{H}_{ij}\in\mathbb{C}^{\widetilde{N}_{i}\times\widetilde{M}_{j}}$.
To achieve a degrees of freedom tuple $\overrightarrow{\boldsymbol{d}}$
for the original system, we need to achieve $T\cdot\overrightarrow{\boldsymbol{d}}$
for this equivalent system, and we can use the exact same precoding
scheme designed for integer-valued corner points.

However, the equivalent channel matrices after symbol-extension are unlike those for their original counterparts (with $T=1$) in that they are block-diagonal.
The primary question that arises is whether the the channel matrices of the time-extended channel continue to yield the linear independence results of the single-letter generic unstructured channels in spite of their special non-generic structure. If they do, then it can be asserted that multi-letter extensions are sufficient to achieve all fractional DoF tuples of $\mathbb{D}_{X}$. 

However, this is not the case in general. Indeed, as it was observed in \cite{jafar2008degrees} the symbol extension technique is not sufficient 
even for the SISO $X$ channel. Interestingly, on the other hand, it is shown in \cite{jafar2008degrees} that, in the symmetric MIMO case, where all nodes have equal
number, $M$, of antennas, and $M>1$, the same idea works. 

Nevertheless, the authors of \cite{jafar2008degrees} claim, based on a few examples, 
that the DoF outer bound region obtained therein is tight in \textquotedblleft most cases\textquotedblright , and give the SISO case as an exception. But it is not clear
if there are other cases that are also such exceptions, and if so, whether they can indeed be seen as exceptions, i.e., it is unclear as to how commonly these exceptions arise, in which just symbol extensions are not enough to achieve all the fractional corner points of $\mathbb{D}_{X}$. This brings us to the next section.

\subsection{Asymmetric Complex Signaling}
\label{sub-section-acs}

As stated previously, since the equivalent channel matrices after symbol-extension
will be block-diagonal, many nice properties of the original generic channels can be lost. It is shown in \cite{jafar2008degrees} that, in the SISO case, the precoding scheme provided previously (with three symbol extensions) fails to achieve the important integer-valued corner point $(1,1,1,1)$ which achieves sum-DoF, because of the block diagonal structure in the extended channel matrices.  

In response to this phenomenon, the authors of \cite{cadambe2010interference} introduced a
new technique named asymmetric complex signaling (ACS). The key idea
of ACS is to allow the inputs to be complex but not circularly symmetric
and use an alternative representation of the channel models in terms
of only real quantities. All dimensions of the new system will be
doubled and all channel coefficients, beamformers, inputs and outputs
will be real-valued. Let $H_{ij}$ $(i,j=1,2)$ be the original complex
channel matrices, their alternative real representations will have
the following forms
\begin{gather}
\hat{H}_{ij}=\left[\begin{array}{cc}
\textrm{Re}(H_{ij})\,\,\,\,\, & -\textrm{Im}(H_{ij})\\
\textrm{Im}(H_{ij})\,\,\,\,\, & \textrm{Re}(H_{ij})
\end{array}\right].\label{eq:hijhead}
\end{gather}
In order to transmit $T\cdot\overrightarrow{\boldsymbol{d}}$ complex-valued
streams over the original system, we need to transmit $2T\cdot\overrightarrow{\boldsymbol{d}}$
real-valued streams over the equivalent real channels.

It is shown in \cite{cadambe2010interference} that using ACS, the
outer bound of $\frac{4}{3}$ degrees of freedom is achievable for
the SISO $X$ channel. In particular, with a three-symbol extension
and ACS, all equivalent channel matrices are of size $6\times6$,
and using the same precoding scheme as used in the other MIMO cases, 
two real-valued symbols can be transmitted via the real channels.
The missing independence requirement in the previous complex-valued
transmission disappears almost surely in this new model. Thus the sum-DoF 
of $4/3$ is achievable (and hence also the DoF region). The readers are referred to \cite{cadambe2010interference} for further details.

\subsection{Closing the gap}

The important question as to whether there are \emph{MIMO} antenna configurations for which, among linear schemes including symbol extensions, ACS is necessary, remains open. The question is also open about whether ACS, along with the other linear techniques, is sufficient for MIMO antenna configurations to achieve all fractional DoF-tuples in $\mathbb{D}_{X}$. If so, for what antennas configurations is it sufficient? Are there DoF-tuples and antenna configurations for which linear precoding schemes including time extensions and ACS are not sufficient?

In this section, all of the above questions are definitively answered. In particular, a class of antenna configurations (that includes the SISO case) are identified that require ACS among linear schemes; i.e., in which just employing symbol extensions alone doesn't suffice. More generally, it is shown that ACS along with the other linear schemes is sufficient to achieve any fractional corner points of the DoF region $\mathbb{D}_{X}$ of the MIMO $X$ channel for any antenna configuration.


\begin{lem}
\label{lem:In-the-case}In the case that $M_{1}+M_{2}=N_{1}+N_{2}$
and $\min(M_{1},M_{2},N_{1},N_{2})=1$, if interference alignment is needed to achieve any
fractional DoF-tuple in $\mathbb{D}_X$, then the achievability scheme
in \ref{sub:Zero-forcing-and-interference}, applied to the $T$-symbol
extended $2 \times 2$ interference network, fails to make the corresponding
symbols distinguishable at the receiver where they are desired. In
particular, if $M_{1}$ or $N_{2}=1$, then $\textrm{span}\left(\widetilde{H}_{21}\widetilde{V}_{21}^{A}\right)\subseteq\textrm{span}\left(\widetilde{H}_{22}\left[\widetilde{V}_{22}^{Z}\ \widetilde{V}_{22}^{A}\right]\right)$;
if $M_{2}$ or $N_{1}=1$, then $\textrm{span}\left(\widetilde{H}_{12}\widetilde{V}_{12}^{A}\right)\subseteq\textrm{span}\left(\widetilde{H}_{11}\left[\widetilde{V}_{11}^{Z}\ \widetilde{V}_{11}^{A}\right]\right)$.\end{lem}
\begin{IEEEproof}
We give the proof of Lemma \ref{lem:In-the-case} in the case that
$M_{1}$ or $N_{2}=1$, and the validity for the case that $M_{2}$
or $N_{1}=1$ follows in the same way.

First, consider the situation when $N_{2}=1$, and we have that $N_{1}=M_{1}+M_{2}-1\geq\max(M_{1},M_{2})$.
Consequently, zero-forcing any symbol of message $W_{21}$ and $W_{22}$
at receiver $R_{1}$ is not possible, i.e., $\widetilde{V}_{22}^{Z}=\widetilde{V}_{21}^{Z}=\emptyset$.
However, since $M_{1}+M_{2}-N_{1}=1$, there exists a one dimensional
null space of the concatenated channel $[H_{11}\,\,H_{12}]$. Thus,
it is possible to align one symbol of message $W_{21}$ with one symbol
of message $W_{22}$ at receiver $R_{1}$. When $T$ channel extensions
are used, the available dimension for interference alignment is
equal to $T$. Suppose the basis vector of the null space of $\mathcal{N}\left([H_{11}\,\,H_{12}]\right)$
is given by\footnote{The dimensions of matrices will be specified in a subscript when such dimensions have to be emphasized or defined for the first time.}
\begin{gather}
\left[\begin{array}{c}
V_{a,\,M_{1}\times1}\\
V_{b,\,M_{2}\times1}
\end{array}\right]_{(M_{1}+M_{2})\times1}\label{eq:vavb}
\end{gather}
 Then one set of basis vectors of $T$-dimensional subspace after
symbol extension will be the column vectors of matrix
\begin{gather}
\left[\begin{array}{cccc}
V_{a,\,M_{1}\times1} & 0 & \ldots & 0\\
0 & V_{a,\,M_{1}\times1} & \ldots & 0\\
\vdots & \vdots & \ddots & \vdots\\
0 & 0 & \ldots & V_{a,\,M_{1}\times1}\\
V_{b,\,M_{2}\times1} & 0 & \ldots & 0\\
0 & V_{b,\,M_{2}\times1} & \ldots & 0\\
\vdots & \vdots & \ddots & \vdots\\
0 & 0 & \ldots & V_{b,\,M_{2}\times1}
\end{array}\right]_{(M_{1}T+M_{2}T)\times T}.\label{eq:vavb1}
\end{gather}
 All beamformers generated from this basis should be of the form
\begin{gather}
[\alpha_{1}V_{a};\,\alpha_{2}V_{a};...;\,\alpha_{T}V_{a};\,\alpha_{1}V_{b};\,\alpha_{2}V_{b};...;\,\alpha_{T}V_{b}],\label{eq:vavb2}
\end{gather}
 where $\alpha_{1},\alpha_{2},...,\alpha_{T}\in\mathbb{C}^{1}$ are
$T$ random scalars. Since $N_{2}=1$, $H_{21}V_{a}$ and $H_{22}V_{b}$
will be scalars. $\widetilde{H}_{21}\widetilde{V}_{21}^{A}$ and $\widetilde{H}_{22}\widetilde{V}_{22}^{A}$
will have the following form
\begin{eqnarray}
\widetilde{H}_{21}\widetilde{V}_{21}^{A} & = & \left[\begin{array}{cccc}
H_{21} & 0 & \ldots & 0\\
0 & H_{21} & \ldots & 0\\
\vdots & \vdots & \ddots & \vdots\\
0 & 0 & \ldots & H_{21}
\end{array}\right]\left[\begin{array}{c}
\alpha_{1}V_{a}\\
\alpha_{2}V_{a}\\
\vdots\\
\alpha_{T}V_{a}
\end{array}\right]\nonumber \\
 & = & \left(H_{21}V_{a}\right)\cdot\left[\begin{array}{c}
\alpha_{1}\\
\alpha_{2}\\
\vdots\\
\alpha_{T}
\end{array}\right],\label{eq:H21A}
\end{eqnarray}
and
\begin{eqnarray}
\widetilde{H}_{22}\widetilde{V}_{22}^{A} & = & \left[\begin{array}{cccc}
H_{22} & 0 & \ldots & 0\\
0 & H_{22} & \ldots & 0\\
\vdots & \vdots & \ddots & \vdots\\
0 & 0 & \ldots & H_{22}
\end{array}\right]\left[\begin{array}{c}
\alpha_{1}V_{b}\\
\alpha_{2}V_{b}\\
\vdots\\
\alpha_{T}V_{b}
\end{array}\right]\nonumber \\
 & = & \left(H_{22}V_{b}\right)\cdot\left[\begin{array}{c}
\alpha_{1}\\
\alpha_{2}\\
\vdots\\
\alpha_{T}
\end{array}\right].\label{eq:H22A}
\end{eqnarray}
Hence $\widetilde{H}_{21}\widetilde{V}_{21}^{A}$ is also aligned
with $\widetilde{H}_{22}\widetilde{V}_{22}^{A}$ at their commonly
desired destination, i.e., $\textrm{span}\left(\widetilde{H}_{21}\widetilde{V}_{21}^{A}\right)=\textrm{span}\left(\widetilde{H}_{22}\widetilde{V}_{22}^{A}\right)$.
This makes $\widetilde{W}_{21}^{A}$ and $\widetilde{W}_{22}^{A}$
indistinguishable at receiver $R_{2}$.

Next, consider the situation when $M_{1}=1$. In this case, $M_{1}\leq N_{1}$
and $M_{2}=N_{1}+N_{2}-1\geq N_{1}$. In other words, the null space
of channel $H_{11}$ does not exist, and the null space of channel
$H_{12}$ may exist. Recall that we only do interference alignment
after zero-forcing of more symbols is not possible. Thus, when interference
alignment is used, $M_{2}-N_{1}$ streams of the message $W_{22}$ 
have already been zero-forced at receiver $R_{1}$. Since $\max(d_{22})=N_{2}>M_{2}-N_{1}$,
it is still possible to transmit another symbol of $W_{22}$. The
dimension of the null-space of the concatenated channel $[H_{11}\,\,H_{12}]$
is equal to $M_{1}+M_{2}-N_{1}$. Letting vector $v_{21}$ be in the subspace
of null space $\mathcal{N}\left(H_{12}\right)$, we have that $\left[\begin{array}{c}
0\\
v_{21}
\end{array}\right]$ will belong to the null space of $\mathcal{N}\left([H_{11}\,\,H_{12}]\right)$.
In other words, $M_{2}-N_{1}$ dimensions
of the null space $\mathcal{N}\left([H_{11}\,\,H_{12}]\right)$ are already occupied 
when doing zero-forcing of message $W_{22}$. The remaining dimension of
null space $\mathcal{N}\left([H_{11}\,\,H_{12}]\right)$ is equal to
$(M_{1}+M_{2}-N_{1})-(M_{2}-N_{1})=M_{1}=1$. Thus, 1 dimension of
interference alignment is possible at receiver $R_{1}$ for messages
$W_{21}$ and $W_{22}$. When $T$ channel extensions are applied,
the available dimension for interference alignment is equal to $T$,
and the dimension of zero-forcing subspace is equal to $T\cdot(M_{2}-N_{1})$.
The beamformers $\widetilde{V}_{21}^{A}$ and $\widetilde{V}_{22}^{A}$
are also in the form of (\ref{eq:vavb})-(\ref{eq:vavb2}). However,
since $N_{2}$ can be greater than 1 in this case, $H_{21}V_{a}$
and $H_{22}V_{b}$ are no longer scalars, and we don't have the desirable
result that $\widetilde{H}_{21}\widetilde{V}_{21}^{A}$ is aligned
with $\widetilde{H}_{22}\widetilde{V}_{22}^{A}$ any more.

To prove that $\textrm{span}\left(\widetilde{H}_{21}\widetilde{V}_{21}^{A}\right)\subseteq\textrm{span}\left(\widetilde{H}_{22}\left[\widetilde{V}_{22}^{Z}\ \widetilde{V}_{22}^{A}\right]\right)$,
we instead prove that $\widetilde{H}_{21}\widetilde{v}_{21,i}^{A}\in\textrm{span}\left(\widetilde{H}_{22}\left[\widetilde{V}_{22}^{Z}\ \widetilde{v}_{22,i}^{A}\right]\right)$,
where $(\widetilde{v}_{21,i}^{A},\widetilde{v}_{22,i}^{A})$ are any
pair of alignment vectors drawn from the same beamformer from the
null space of $\mathcal{N}\left([H_{11}\,\,H_{12}]\right)$. Then,
we will have that $\underset{i}{\cup}\textrm{span}\left(\widetilde{H}_{21}\widetilde{v}_{21,i}^{A}\right)\subseteq\underset{i}{\cup}\textrm{span}\left(\widetilde{H}_{22}\left[\widetilde{V}_{22}^{Z}\ \widetilde{v}_{22,i}^{A}\right]\right)$,
which is the desired result. Let $\widetilde{v}_{21,i}^{A}=[\alpha_{1}V_{a};\,\alpha_{2}V_{a};...;\,\alpha_{T}V_{a}]$
and $\widetilde{v}_{22,i}^{A}=[\alpha_{1}V_{b};\,\alpha_{2}V_{b};...;\,\alpha_{T}V_{b}]$,
we have that
\begin{gather*}
\widetilde{H}_{21}\widetilde{V}_{21}^{A}=\left[\begin{array}{cccc}
H_{21}V_{a} & 0 & \ldots & 0\\
0 & H_{21}V_{a} & \ldots & 0\\
\vdots & \vdots & \ddots & \vdots\\
0 & 0 & \ldots & H_{21}V_{a}
\end{array}\right]\left[\begin{array}{c}
\alpha_{1}\\
\alpha_{2}\\
\vdots\\
\alpha_{T}
\end{array}\right]
\end{gather*}
and
\begin{gather*}
\widetilde{H}_{22}\widetilde{V}_{22}^{A}=\left[\begin{array}{cccc}
H_{22}V_{b} & 0 & \ldots & 0\\
0 & H_{22}V_{b} & \ldots & 0\\
\vdots & \vdots & \ddots & \vdots\\
0 & 0 & \ldots & H_{22}V_{b}
\end{array}\right]\left[\begin{array}{c}
\alpha_{1}\\
\alpha_{2}\\
\vdots\\
\alpha_{T}
\end{array}\right].
\end{gather*}
Let column vectors of $V_{22}^{Z}$ be a basis of the nullspace $\mathcal{N}\left(H_{12}\right)$.
We have that
\begin{gather*}
\widetilde{H}_{22}\widetilde{V}_{22}^{Z}=\left[\begin{array}{cccc}
H_{22}V_{22}^{Z} & 0 & \ldots & 0\\
0 & H_{22}V_{22}^{Z} & \ldots & 0\\
\vdots & \vdots & \ddots & \vdots\\
0 & 0 & \ldots & H_{22}V_{22}^{Z}
\end{array}\right].
\end{gather*}
Note that the $T\cdot N_{2}$ dimensional space at receiver
$R_{2}$ can be partitioned into $T$ linearly independent subspaces
according to different symbol extension slot index. In order to prove
that $\widetilde{H}_{21}\widetilde{v}_{21,i}^{A}\in\textrm{span}\left(\widetilde{H}_{22}\left[\widetilde{V}_{22}^{Z}\ \widetilde{v}_{22,i}^{A}\right]\right)$,
it is sufficient to prove that $H_{21}V_{a}\alpha_{i}\in\textrm{span}\left(H_{22}[V_{22}^{Z}\ V_{b}\alpha_{i}]\right)$
for all $i=1,...,T$. Since $\alpha_{i}$ here are all scalars,
we only need to show that $H_{21}V_{a}\in\textrm{span}\left(H_{22}[V_{22}^{Z}\ V_{b}]\right)$. 

Because$V_{22}^{Z}$ is generated from the nullspace of channel $H_{12}$,
it is independent with channel matrix $H_{22}$. Consequently, $H_{22}V_{22}^{Z}$
will almost surely reserve the column rank of $V_{22}^{Z}$, since
$H_{22}$ is a generic full matrix whose rank is greater than $V_{22}^{Z}$'s.
In other words, $\textrm{rank}(H_{22}V_{22}^{Z})=M_{2}-N_{1}=N_{2}-1$
almost surely. Since $V_{b}$ is linearly independent with the column
vectors of $V_{22}^{Z}$, $H_{22}V_{b}$ will also be linear independent
with the column vectors of $H_{22}V_{22}^{Z}$ almost surely. Thus,
$\textrm{rank}(H_{22}[V_{22}^{Z}\ V_{b}])=(N_{2}-1)+1=N_{2}$. In
other words, the column vectors of $H_{22}[V_{22}^{Z}\ V_{b}]$ would
span the entire $N_{2}$-dimensional subspace at receiver $R_{2}$.
Since vector $H_{21}V_{a}$ also belongs to the same subspace, we
have that $H_{21}V_{a}\in\textrm{span}\left(H_{22}[V_{22}^{Z}\ V_{b}]\right)$,
which leads to that $\widetilde{H}_{21}\widetilde{v}_{21,i}^{A}\in\textrm{span}\left(\widetilde{H}_{22}\left[\widetilde{V}_{22}^{Z}\ \widetilde{v}_{22,i}^{A}\right]\right)$.
Thus, message $\widetilde{W}_{21}^{A}$ and $\widetilde{W}_{22}^{A}$
are indistinguishable at receiver $R_{2}$.\end{IEEEproof}
\begin{lem}
By using the technique of ACS together with symbol extensions,
the problem of unexpected alignment of desired messages is avoided.\end{lem}
\begin{IEEEproof}
The equivalent channel matrices, when doing $T$-symbol extension
and ACS, are given as $\bar{H}_{ij}=I_{T\times T}\otimes\hat{H}_{ij},$
where $\hat{H}_{ij}$ is given in equation \ref{eq:hijhead}. We need
to transmit $2T\cdot\overrightarrow{\boldsymbol{d}}$ real-valued
streams over the equivalent real channels.

Consider again the independence of $\bar{H}_{21}\bar{V}_{21}^{A}$
and $\bar{H}_{22}\bar{V}_{22}^{A}$ for the cases in Lemma \ref{lem:In-the-case}.
If $N_{2}=1$, when doing asymmetric complex signaling, the dimension
of $V_{a}$ and $V_{b}$ in (\ref{eq:vavb}) will be $2M_{1}\times2$
and $2M_{2}\times2$, respectively. $\bar{H}_{21}\bar{V}_{21}^{A}$
and $\bar{H}_{22}\bar{V}_{22}^{A}$ will instead have the following
form 
\begin{eqnarray*}
\bar{H}_{21}\bar{V}_{21}^{A} & = & \left[\begin{array}{cccc}
\hat{H}_{21}V_{a} & 0 & \ldots & 0\\
0 & \hat{H}_{21}V_{a} & \ldots & 0\\
\vdots & \vdots & \ddots & \vdots\\
0 & 0 & \ldots & \hat{H}_{21}V_{a}
\end{array}\right]\left[\begin{array}{c}
\alpha_{1}\\
\alpha_{2}\\
\vdots\\
\alpha_{2T}
\end{array}\right]
\end{eqnarray*}
and
\begin{eqnarray*}
\bar{H}_{22}\bar{V}_{22}^{A} & = & \left[\begin{array}{cccc}
\hat{H}_{22}V_{b} & 0 & \ldots & 0\\
0 & \hat{H}_{22}V_{b} & \ldots & 0\\
\vdots & \vdots & \ddots & \vdots\\
0 & 0 & \ldots & \hat{H}_{22}V_{b}
\end{array}\right]\left[\begin{array}{c}
\alpha_{1}\\
\alpha_{2}\\
\vdots\\
\alpha_{2T}
\end{array}\right].
\end{eqnarray*}
where $\alpha_{1},\alpha_{2},...,\alpha_{2T}\in\mathbb{R}^{1}$ are
$2T$ random real scalars. Now, $\hat{H}_{21}V_{a}$ and $\hat{H}_{22}V_{b}$
are both $2\times2$ real matrices rather than scalars as in (\ref{eq:H21A})
and (\ref{eq:H22A}). Each diagonal block of $\hat{H}_{21}V_{a}$
or $\hat{H}_{22}V_{b}$ works as if it is to rotate a random $2\times1$
real vector with a certain degree. However, the randomness of $[\alpha_{1;}\ ...;\ \alpha_{2T}]$
makes the projections in different symbol extension slots independent
with each other. Thus, $\bar{H}_{21}\bar{V}_{21}^{A}$ and $\bar{H}_{22}\bar{V}_{22}^{A}$
will be linearly independent almost surely.

Consider again the case that $M_{1}=1$, the column rank of $\hat{H}_{22}[V_{22}^{Z}\ V_{b}]$
will be equal to $2(N_{2}-1)+1=N_{2}-1$. In other words, there is
still 1 dimension left in the receiver subspace. Thus, $\hat{H}_{21}V_{a}$
is independent with $\hat{H}_{22}[V_{22}^{Z}\ V_{b}]$ almost surely.
In the situation that the column dimension of $V_{a}$ and $V_{b}$
are $n$, which is greater than 1, i.e., there are multiple pairs
of symbols to be aligned, the column size of $\left[\hat{H}_{22}[V_{22}^{Z}\ V_{b}]\ \hat{H}_{21}V_{a}\right]$
will be equal to $2(N_{2}-1)+n+n>N_{2}$ and thus the columns
of $\hat{H}_{22}[V_{22}^{Z}\ V_{b}]$ are linear dependent with columns
of $\hat{H}_{21}V_{a}$. However, since the coefficients required for
dependence for $\hat{H}_{21}V_{a}$ are different almost surely in
different time slots, the $2\cdot T$ dimensional $\bar{H}_{22}[\bar{V}_{22}^{Z}\ \bar{V}_{22}^{A}]$
will still be linearly independent with $\bar{H}_{12}\bar{V}_{12}^{A}$
almost surely, so long as $d_{22}^{Z}+d_{22}^{A}+d_{12}^{A}\leq T\cdot N_{2}$. 

In summary, the desired messages are still linearly independent with each
other at both receiver.\end{IEEEproof}
\begin{rem}
The authors of \cite{cadambe2010interference} introduced ACS in the context of the SISO $X$ channel and showed that the total DoF of $\frac{4}{3}$ can be achieved in that channel. In this paper, we provided
a new and simplified perspective on how ACS works. In particular,
it transforms the previous scalar multiplication to a local vector rotation,
thus obviating the unexpected linear dependences among all the beamformers.
Using this we broaden its applicability to MIMO $X$ channel, and more generally, ia a later section, to the nine-message MIMO 2$\times$2 interference network.
\end{rem}

\begin{rem}
For all the other antenna settings not included in the cases given
in Lemma \ref{lem:In-the-case}, there is no unexpected loss of independence
of desired messages when doing symbol extensions. Thus, ACS is not necessary in those cases.
\end{rem}

\subsection{Further results on the MIMO $X$ channel}

In this section, we discuss several other observations/results about
the MIMO $X$ channel.
\begin{lem}
\label{lem:In-the-symmetric}In the symmetric $(M,M,N,N)$ antenna
setting, the maximum sum DoF of the MIMO $X$ channel is given by
\begin{gather*}
\begin{cases}
\begin{array}{cccc}
2M, & \textrm{if} &  & 0<\frac{M}{N}\leq\frac{2}{3}\\
\frac{4N}{3}, & \textrm{if} &  & \frac{2}{3}<\frac{M}{N}\leq1\\
\frac{4M}{3}, & \textrm{if} &  & 1<\frac{M}{N}\leq\frac{3}{2}\\
2N, & \textrm{if} &  & \frac{3}{2}<\frac{M}{N}
\end{array} & .\end{cases}
\end{gather*}

\end{lem}
Lemma \ref{lem:In-the-symmetric} is a special case of Theorem \ref{thm-mimo-x}.
In terms of sum-DoF performance, there are hence redundant antennas at the
transmitters if $\frac{2}{3}<\frac{M}{N}\leq1$ or $\frac{M}{N}>\frac{3}{2}$,
and there are redundant antennas at the receivers if $0<\frac{M}{N}<\frac{2}{3}$
or $1\leq\frac{M}{N}<\frac{3}{2}$. In the case that $M=N$, the redundancy
exists both at the transmitters and at the receivers. For example,
the three antenna settings of $(3,3,3,3)$, $(3,3,2,2)$ and $(2,2,3,3)$
all have the same maximum sum-DoF of 4. 

Interestingly, for the equal-antenna case of $(3,3,3,3)$, one
can easily achieve the DoF-tuple of $(1,1,1,1)$ by turning off
one antenna at each receiver and then transmitting all four symbols of the
private messages using zero-forcing beamforming in each of the one-dimensional null space of the remaining channel matrices. No explicit interference alignment is actually needed to achieve the optimal sum-DoF. Given that explicit interference alignment was first discovered in the context of the symmetric three-antenna MIMO $X$ channel as being the key ingredient \cite{jafar2008degrees} needed to achieve DoF-optimality, this observation is surprising. To the best of the authors' knowledge, this is the first time this simple result has been noted. Shutting down the redundant antenna at each receiver could however be seen as implicitly aligning interference in a subspace that would only be seen by that antenna and then discarding that subspace.
\begin{lem}
For the special cases given in Lemma \ref{lem:In-the-case}, in which
ACS is required to achieve the maximum sum-DoF, the maximum sum-DoF is equal to $C-\frac{2}{3}$, where $C=M_{1}+M_{2}=N_{1}+N_{2}$.
The DoF tuple to achieve the maximum sum-DoF is given by $d_{ij}=\min(M_{j},N_{i})-\frac{2}{3}$.\end{lem}
\begin{IEEEproof}
We give the proof for the case that $M_{1}=\min(M_{1},M_{2},N_{1},N_{2})=1$
here. The other cases follow in the same way.

Since $M_{1}=1$ and $M_{1}+M_{2}=N_{1}+N_{2}=C$, we have that $\max(M_{1},N_{1})=N_{1}$,
$\max(M_{1},N_{2})=N_{2}$, $\max(M_{2},N_{1})=M_{2}$ and $\max(M_{2},N_{1})=M_{2}$.
Adding the first 4 inequalities in $\mathbb{D}_{X}$ together, we
have that 
\begin{gather*}
3(d_{11}+d_{12}+d_{21}+d_{22})\leq N_{1}+N_{2}+2M_{2}=3C-2
\end{gather*}
Thus, the sum-DoF is bounded by $C-\frac{2}{3}$. It is easy to verify
that the DoF tuple $d_{ij}=\min(M_{j},N_{i})-\frac{2}{3}$ $(i,j=1,2)$
achieves the optimal sum DoF and is within the DoF region $\mathbb{D}_{X}$
. Thus, the maximum sum DoF is equal to $C-\frac{2}{3}$.

A symbol extension of length 3, together with ACS, is required to
achieve this corner point.\end{IEEEproof}
\begin{lem}
In the case that only three private messages are transmitted in the
channel, all the corner points will be integer-valued. Thus, neither
symbol extension nor ACS is necessary.\end{lem}
\begin{IEEEproof}
Since the channel is isotropic with respect to any message, we can
assume without loss of generality that the three private messages
are $W_{11}$, $W_{12}$ and $W_{21}$. By deleting $d_{22}$ from
$\mathbb{D}_{X}$ and removing the redundant inequalities, we obtain
the following 3-dimensional DoF region.
\begin{align}
\mathbb{D}^{'}=\lbrace & (d_{11},d_{21},d_{12})\in\mathbb{R}_{+}^{3}:\nonumber \\
 & d_{11}+d_{12}+d_{21}\leq\max(M_{1},N_{1}),\label{eq:3d1}\\
 & d_{11}+d_{12}\leq N_{1},\label{eq:3d2}\\
 & d_{21}+d_{11}\leq M_{1}\label{eq:3d3}\\
 & d_{21}+d_{12}\leq\max(M_{2},N_{2}),\label{eq:3d4}\\
 & d_{21}\leq N_{2},\label{eq:3d5}\\
 & d_{12}\leq M_{2}\rbrace.\label{eq:3d6}
\end{align}
Each corner point of this 3-D region will be the intersection of three
of the nine facets describing the polytope. Observing the constraints,
it is easy to verify that the only possible combination of facets
that can have a fractional intersection are (\ref{eq:3d2}), (\ref{eq:3d3})
and (\ref{eq:3d4}), and the corresponding vertex is
\begin{gather*}
\begin{cases}
\begin{array}{c}
d_{11}=\frac{M_{1}+N_{1}-\max(M_{2},N_{2})}{2}\\
d_{12}=\frac{N_{1}+\max(M_{2},N_{2})-M_{1}}{2}\\
d_{21}=\frac{M_{1}+\max(M_{2},N_{2})-N_{1}}{2}
\end{array} & .\end{cases}
\end{gather*}
These three values will be all integers or all non-integer fractions which are an odd-multiple of $\frac{1}{2}$.

From constraint (\ref{eq:3d1}), we have that
\begin{gather*}
\frac{M_{1}+N_{1}+\max(M_{2},N_{2})}{2}\leq\max(M_{1},N_{1}).
\end{gather*}
Otherwise, this corner point will be outside the DoF region. Consequently,
one of $d_{12}$ and $d_{21}$ will be $ \leq 0$. If
it is less than zero, this corner point is outside the DoF region and
therefore irrelevant; if it is equal to 0, then the other two values will 
be integers.

The intersection of all other combinations of facets will be integer-valued,
thus, all the corner points of $\mathbb{D}^{'}$ are integer-valued,
and neither symbol extension nor ACS are
necessary to achieve them.\end{IEEEproof}
\begin{lem}
For the MIMO $X$ channel of an arbitrary antenna setting, if there are
fractional-valued corner points and symbol extension is required
to achieve this corner point, the length of symbol extension will
be at most 3. \end{lem}
\begin{IEEEproof}
Again, each corner point of the 4-dimensional DoF region is the intersection
of four of the facets describing the polytope. Since the coefficients
of any facet are either 0 or 1, any selected 4-by-4 coefficient matrix
will be a binary matrix. According to the Hadamard maximal determinant
problem \cite{Brenner1972}, the determinant of an order 4 binary
matrix can at most be 3. Consequently, the inverse of any 4-by-4 coefficient
matrix, if it exists, can at most have a denominator of 3. Thus, for any 
non-integer valued corner points, the denominator
will be at most 3. Thus, the length of symbol extension will be at
most 3.

More specifically in this problem, it is shown that there is only
one corner point whose denominator is 3, and this corner point is
the intersection of the four facets corresponding to the first four
constraints in $\mathbb{D}_{X}$.
\end{IEEEproof}

\subsection{Cognitive MIMO $X$ channel}

If one of the four private messages in the MIMO $X$ channel, for
example $W_{11}$, is made available non-causally at the other transmitter,
the channel is named cognitive MIMO $X$ channel. It is shown in \cite{jafar2008degrees}
that the sum DoF of the cognitive MIMO $X$ channel with equal number,
$M$, of antennas at each terminal is equal to $\frac{3}{2}M$, which
is greater than the sum DoF of $\frac{4}{3}M$ of the symmetric $X$
channel. So, cognitive message sharing helps increase sum DoF in this
case. We discuss more general properties of the cognitive MIMO $X$ channel here.
\begin{thm}
\label{thm:The-degrees-of-cogX}The degrees of freedom region of the
cognitive MIMO $X$ channel with message $W_{21}$, $W_{12}$, $W_{22}$
and $W_{01}$ is given by 
\begin{align*}
\mathbb{D}_{co-X}=\bigl\{ & (d_{01},d_{21},d_{12},d_{22})\in\mathbb{R}_{+}^{4}:\\
 & d_{01}+d_{12}+d_{21}\leq\max(M_{1},N_{1}),\\
 & d_{01}+d_{12}+d_{22}\leq\max(M_{2},N_{1}),\\
 & d_{21}+d_{22}+d_{12}\leq\max(M_{2},N_{2}),\\
 & d_{01}+d_{12}\leq N_{1},\,\,d_{21}+d_{22}\leq N_{2},\\
 & d_{21}\leq M_{1},\,\,d_{12}+d_{22}\leq M_{2},\\
 & d_{01}+d_{21}+d_{12}+d_{22}\leq M_{1}+M_{2}\bigr\}
\end{align*}

\end{thm}
Theorem \ref{thm:The-degrees-of-cogX} follows directly from our main
result of the 9-dimensional DoF region of the MIMO $2 \times 2$  Gaussian
interference network with general message sets given in Section \ref{sec:Main-Result}.
When the above DoF {\em region} is specialized to the symmetric,
equal-antenna case, all but the first three bounds are redundant,
and it is easy to see that the DoF-tuple $(d_{01}=M/2, d_{12}=0, d_{21}=M/2, d_{22}=M/2)$,
the achievability of which was shown in \cite{jafar2008degrees} for
$M>1$ (using two-symbol extensions), is a maximum sum-DoF corner
point of $\mathbb{D}_{co-X}$ for any $M\geq 1$. 

More generally, the DoF region of cognitive MIMO $X$ channel is in general
greater than that of the MIMO $X$ channel. For example, consider the case
of $M_{1}=3$, $M_{2}=4$, $N_{1}=5$, $N_{2}=6$. When $d_{12}$,
$d_{21}$ and $d_{22}$ are all set to be equal to 1, $d_{11} $ can
be at most 2 in the MIMO $X$ channel, whereas $d_{01} $ can be up to 3 in
the cognitive MIMO $X$ channel. Even the cognition of one message among
the transmitters can significantly improve the maximum achievable DoF.
\begin{lem}
\label{lem:In-the-symmetric-1}In the symmetric $(M,M,N,N)$ antenna
setting, the maximum sum DoF of the cognitive MIMO $X$ channel is
given by
\begin{gather*}
\begin{cases}
\begin{array}{cccc}
2M, & \textrm{if} &  & 0<\frac{M}{N}\leq\frac{3}{4}\\
\frac{3N}{2}, & \textrm{if} &  & \frac{3}{4}<\frac{M}{N}\leq1\\
M+\frac{N}{2}, & \textrm{if} &  & 1<\frac{M}{N}\leq\frac{3}{2}\\
2N, & \textrm{if} &  & \frac{3}{2}<\frac{M}{N}
\end{array} & .\end{cases}
\end{gather*}

\end{lem}
Lemma \ref{lem:In-the-symmetric-1} is a special case of Theorem \ref{thm:The-degrees-of-cogX}. Comparing with the result of MIMO $X$ channel,
the sum DoF of the cognitive MIMO $X$ is strictly greater than that
of the MIMO $X$ when $\frac{2}{3}<\frac{M}{N}<\frac{3}{2}$. When $ \frac{M}{N} \leq \frac{2}{3}$ or $ \frac{M}{N} \geq \frac{3}{2}$, there are redundant antennas at the transmitters or the receivers, and message cognition does not help in improving the sum
DoF of the system.
\begin{lem}
\label{lem:In-the-case-1}In the case that $M_{1}+M_{2}=N_{1}+N_{2}$
and $\min(M_{1},M_{2},N_{1},N_{2})=1$, among linear strategies, ACS
is required to achieve the DoF region of the cognitive MIMO $X$ channel.\end{lem}
\begin{IEEEproof}
The reason that ACS is necessary for the cognitive
MIMO $X$ channel is the same as that for the MIMO $X$ channel
in lemma \ref{lem:In-the-case}. We omit the details for brevity\end{IEEEproof}
\begin{lem}
For the special cases given in Lemma \ref{lem:In-the-case-1}, in
which ACS is required to achieve the maximum
sum-DoF of the cognitive MIMO $X$ channel, the maximum sum-DoF is
equal to $C-\frac{1}{2}$, where $C=M_{1}+M_{2}=N_{1}+N_{2}$. The
DoF tuple to achieve the maximum sum-DoF is given by $(d_{01},d_{21},d_{12},d_{22})$
=$(\min(M_{1},N_{1})-\frac{1}{2},\ \min(M_{1},N_{2})-\frac{1}{2},\ \min(M_{1}+M_{2},N_{1})-\min(M_{1},N_{1}),\ \min(M_{2},N_{2})-\frac{1}{2})$
or $(\min(M_{1}+M_{2},N_{1})-\frac{1}{2},\ \min(M_{1},N_{2})-\frac{1}{2},\ 0,\ \min(M_{2},N_{2})-\frac{1}{2})$.\end{lem}
\begin{IEEEproof}
Adding the 1st, 2nd and 5th inequalities in $\mathbb{D}_{co-X}$  together,
we have that $2d_{sum}\leq\max(M_{1},N_{1})+\max(M_{2},N_{1})+N_{2}$,
which is always equal to $2C-1$. Thus, the sum DoF is upper bounded
by $C-\frac{1}{2}$. One can easily verify that the two given DoF
tuples are both within $\mathbb{D}_{co-X}$ and achieve the maximum
sum-DoF.

A symbol extension of length 2, together with ACS, is required to
achieve this corner point.
\end{IEEEproof}
There can be two non-integer-valued corner points which achieve the
maximum sum-DoF. However, when $\textrm{min}(M_{1}+M_{2},N_{1})=\textrm{min}(M_{1},N_{1})$,
or equivalently $M_{1}\geq N_{1}$, these two corner points are the
same. If these two corner points are different, we can get one of
them from the other by just regarding the non-zero $d_{12}$ symbols
of message $W_{12}$ as part of message $W_{01}$.
\begin{lem}
\label{lem:For-the-cognitive2}For the cognitive MIMO $X$ channel
of arbitrary antenna setting, if there are any fractional-valued corner
point and symbol extensions are required to achieve this corner point,
the length of symbol extension will be at most 2. \end{lem}
\begin{IEEEproof}
Although the determinant of an arbitrary 4-by-4 binary matrix can be at most 3, it is easy to verify that the maximum determinant of any 4-by-4 coefficient matrix generating from any four facets given in $\mathbb{D}_{co-X}$
  is equal to 2. Thus, the length of symbol extension will be at most 2.
\end{IEEEproof}

\section{\label{sec:Main-Result}Main Result}

Now, let us consider the general MIMO $2 \times 2$ interference network
with nine messages.

The following theorem gives the nine-dimensional DoF
region of the MIMO $2 \times 2$  Gaussian interference network with
general message sets.
\begin{thm}
\label{thm:The-total-degrees}The degrees of freedom region of the
MIMO $2 \times 2$ Gaussian interference network with the general message
set is $\mathbb{D}=$
\begin{align}
\mathbb{} & \lbrace(d_{11},d_{21},d_{12},d_{22},d_{1},d_{2},d_{01},d_{02},d_{0})\in\mathbb{R}_{+}^{\mathsf{E}}:\nonumber \\
 & d_{1}+d_{2}+d_{0}+d_{01}+d_{11}+d_{12}+d_{21}\leq\max(M_{1},N_{1})\label{eq:out1}\\
 & d_{1}+d_{2}+d_{0}+d_{01}+d_{11}+d_{12}+d_{22}\leq\max(M_{2},N_{1})\label{eq:out2}\\
 & d_{1}+d_{2}+d_{0}+d_{02}+d_{21}+d_{22}+d_{11}\leq\max(M_{1},N_{2})\label{eq:out3}\\
 & d_{1}+d_{2}+d_{0}+d_{02}+d_{21}+d_{22}+d_{12}\leq\max(M_{2},N_{2})\label{eq:out4}\\
 & d_{1}+d_{2}+d_{0}+d_{01}+d_{11}+d_{12}\leq N_{1}\label{eq:out5}\\
 & d_{1}+d_{2}+d_{0}+d_{02}+d_{21}+d_{22}\leq N_{2}\label{eq:out6}\\
 & d_{1}+d_{11}+d_{21}\leq M_{1}\label{eq:out7}\\
 & d_{2}+d_{12}+d_{22}\leq M_{2}\label{eq:out8}\\
 & d_{1}+d_{2}+d_{0}+d_{01}+d_{02}+d_{11}+d_{21}+d_{12}+d_{22}\nonumber \\
 & \;\;\;\leq\min(M_{1}+M_{2},N_{1}+N_{2})\rbrace,\label{eq:out9}
\end{align}
\end{thm}
\begin{IEEEproof}
The proof of $\mathbb{D}$ being an outer bound is given in Section
\ref{sec:Outerbound-on-the}. The inner bound is given in the Lemmas
\ref{lem:An-inner-bound} and \ref{lem:The-fractional-numbers} in
this section.\end{IEEEproof}
\begin{lem}
\label{lem:An-inner-bound}An inner bound to the degrees of freedom
region of the MIMO  $2 \times 2$ interference network with general
message set is \textup{$\mathbb{D}_{\mathrm{in}}=\textrm{co}\left(\mathbb{D}\cap\mathbb{Z}_{+}^{9}\right)$,}
i.e., all the integer-valued degrees of freedom in $\mathbb{D}$ as
well as their convex hull are achievable.
\end{lem}
\textit{Outline of Proof:} In this outline, we will describe a method
to construct the transmit beamformers for various messages. It will
be shown later in Section \ref{sec:Achievability-of-the} that using
this scheme the DoF region $\mathbb{D}_{\mathrm{in}}$ can be achieved.

To achieve any integer-valued nine-dimensional DoF tuple $\boldsymbol{\overrightarrow{d}}=(d_{11},d_{21},d_{12},d_{22},d_{1},d_{2},d_{01},d_{02},d_{0})$
within $\mathbb{D}$, we use the following precoding scheme.

Consider linear beamforming. Expressing received signals at receive
$r$ ($r$=1,2) in the form of different messages, we have 
\begin{eqnarray*}
Y_{r} & = & H_{r1}\cdot\left(V_{11}S_{11}+V_{21}S_{21}+V_{1}S_{1}\right)\\
 &  & +H_{r2}\cdot\left(V_{12}S_{12}+V_{22}S_{22}+V_{2}S_{2}\right)\\
 &  & +\left[H_{r1}\,\,\,H_{r2}\right]\cdot\left(V_{01}S_{01}+V_{02}S_{02}+V_{0}S_{0}\right)+Z_{r},
\end{eqnarray*}
 where $S_x$ and $V_x$ denote the symbols and the corresponding precoding matrices for the message with index $x\in\mathsf{E}$. 
Let the column size of $V_{x}$ is equal to $d_{x}$.

\begin{table}
\caption{\label{tab:Message-Grouping-and}Message Grouping and Corresponding
Precoding Methods}

\centering{}%
\begin{tabular}{|>{\raggedright}m{0.95cm}|>{\raggedright}m{7cm}|}
\hline 
Group 1 & ($W_{11}$, $W_{12}$, $W_{21}$ and $W_{22}$)

In $W_{11}$, for example, there are $d_{11}$ independent symbols.
$d_{11}^{Z}$ of them are zero-forced at receiver $R_{2}$, $d_{11}^{A}$
of them are aligned with part of $W_{12}$ at receiver $R_{2}$. The
remaining $d_{11}^{R}=d_{11}-d_{11}^{Z}-d_{11}^{A}$ symbols are transmitted
using random beamforming.\tabularnewline
\hline 
Group 2 & ($W_{01}$ and $W_{02}$)

In $W_{01}$ for example, there are $d_{01}$ independent symbols.
$d_{01}^{Z}$ of them are zero-forced at receiver $R_{2}$, and the
remaining $d_{01}^{R}=d_{01}-d_{01}^{Z}$ symbols are transmitted
using random beamforming.\tabularnewline
\hline 
Group 3 & ($W_{1}$, $W_{2}$ and $W_{0}$)

Random beamforming is used for all symbols of this group.\tabularnewline
\hline 
\end{tabular}
\end{table}

The techniques used here are transmit zero-forcing, interference
alignment and random beamforming. 

The nine messages are divided into three groups as shown in Table 
\ref{tab:Message-Grouping-and}.
Group 1 consists of the four point-to-point private or $X$-channel 
messages \{$W_{11}$,
$W_{12}$, $W_{21}$, $W_{22}$\}, Group 2 consists of the cognitive
and common messages which are known to both transmitters, namely, \{$W_{01}$,
$W_{02}$\}. Group 3 consists of the remaining three multicast
messages \{$W_{1}$, $W_{2}$, $W_{0}$\}. The transmission of Group
1 messages is done in the exact same way as in the MIMO $X$ channel.
Then, the other two groups are transmitted via the channel resources
still available. Recall that, for Group 1, message $W_{ij}$ $(i,j=1,2)$
is partitioned into three linearly independent parts, i.e., $W_{ij}^{Z}$,
$W_{ij}^{A}$ and $W_{ij}^{R}$. Here, for Group 2, message $W_{0i}$
$(i=1,2)$ is partitioned into two linearly independent parts, namely,
$W_{0i}^{Z}$ and $W_{0i}^{R}$. For Group 3, all messages are transmitted
using random beamforming, since no interference elimination is necessary
for them. Thus, message $W_{k}$ $(k=0,1,2)$ is all classified as
$W_{k}^{R}$. We have that
\begin{align*}
 & d_{ij}=d_{ij}^{Z}+d_{ij}^{A}+d_{ij}^{R}\\
 & d_{0i}=d_{0i}^{Z}+d_{0i}^{R}\\
 & d_{k}=d_{k}^{R}
\end{align*}
and
\begin{align*}
 & V_{ij}=[V_{ij}^{Z}\ V_{ij}^{A}\ V_{ij}^{R}]\\
 & V_{0i}=[V_{0i}^{Z}\ V_{0i}^{R}]\\
 & V_{k}=V_{k}^{R}
\end{align*}
where $i,j=1,2$ and $k=0,1,2$. The dimensions of different parts
of each message are given as follows
\begin{align}
 & d_{ij}^{Z}=\min\bigl(d_{ij},(M_{j}-N_{\widehat{i}})^{+}\bigr)\label{eq:ddd1}\\
 & d_{i1}^{A}=d_{i2}^{A}=\min\bigl(d_{i1}-d_{i1}^{Z},d_{i2}-d_{i2}^{Z},\nonumber \\
 & \quad\quad\quad\quad(M_{1}+M_{2}-N_{\widehat{i}}-d_{i1}^{Z}-d_{i2}^{Z})^{+}\bigr)\label{eq:ddd2}\\
 & d_{ij}^{R}=d_{ij}-d_{ij}^{Z}-d_{ij}^{A}\label{eq:ddd3}\\
 & d_{0i}^{Z}=\min\bigl(d_{0i},(M_{1}+M_{2}-N_{\widehat{i}}-d_{i1}^{Z}-d_{i2}^{Z}-d_{i1}^{A})^{+}\bigr)\nonumber \\
\label{eq:ddd4}\\
 & d_{0i}^{R}=d_{0i}-d_{0i}^{Z}\label{eq:ddd5}\\
 & d_{k}^{R}=d_{k},\label{eq:ddd6}
\end{align}
where $i,j=1,2$, $\widehat{i}=3-i$, and $k=0,1,2.$ To make the
expressions more succinct, we define following auxiliary variables:
\begin{eqnarray}
Z_{ij} & \equiv & d_{ij}^{Z}\label{eq:z_a_d_1}\\
A_{i} & \equiv & d_{i1}^{A}=d_{i2}^{A}\label{eq:z_a_d_2}\\
Z_{0i} & \equiv & d_{0i}^{Z}.\label{eq:z_a_d_3}
\end{eqnarray}
These values are pre-determined according to the value of the DoF
tuple and the system antenna setting. They naturally follow from the
fact that the numbers of beamformers transmitted using zero-forcing
or interference alignment cannot exceed the corresponding available
null space dimensions. For the four private messages, if zero-forcing
is possible, use zero-forcing first. If there are 
more streams that must be send, use interference alignment next. If
there are still more streams after running out of the possibility of
doing alignment, use random beamforming. For the two cognitive and
common messages, if there are residual available null space dimensions,
transmit using zero-forcing; otherwise, just use random beamforming.
For three multicast messages, all streams are transmitted using random
beamforming. 

The key to using zero-forcing or interference alignment is to appropriately
utilize the beamformers picking from the null space of corresponding
channels. For a generic channel matrix $H_{n\times m}$ $(n<m)$,
the dimension of its nullspace is equal to $m-n$. To obtain a basis
of $\mathcal{N}(H)$, we can do a singular value decomposition (SVD)
of matrix $H$ while arranging the singular values in non-increasing
order. Then, the last $m-n$ right-singular column vectors, which
are corresponding to singular value 0, will form a basis of $\mathcal{N}(H)$.
We construct matrix $\varPhi(H)$ such that its column vectors are
equal to these basis vectors of $\mathcal{N}(H)$. Let matrix $X_{(m-n)\times a}$
denote a randomly $(m-n)\times a$ matrix, whose column vectors are
generated independently from a uniform distribution on a $m-n$ dimensional
sphere of radius 1. Then, $\varPhi(H)\cdot X_{(m-n)\times b}$ will
generate $b$ random combinations of these basis vectors. If $b\leq m-n$,
these $b$ vectors will be linearly independent of each other almost
surely.

Now, construct the beamformers for all 9 messages according to the
equations (\ref{eq:vvv1})-(\ref{eq:vvv7}) listed below.
\begin{eqnarray}
V_{ij}^{Z} & = & \varPhi(H_{\widehat{i}j})\cdot X_{ij,(M_{j}-N_{\widehat{i}})\times d_{ij}^{Z}}^{Z}\label{eq:vvv1}\\
\left[\begin{array}{c}
V_{i1}^{A}\\
V_{i2}^{A}
\end{array}\right] & = & \varPhi([H_{\widehat{i}1}\ H_{\widehat{i}2}])\cdot X_{i,(M_{1}+M_{2}-N_{\widehat{i}})\times d_{i1}^{A}}^{A}\label{eq:vvv2}\\
V_{ij}^{R} & = & X_{ij,M_{j}\times d_{ij}^{R}}^{R}\label{eq:vvv3}\\
V_{0i}^{Z} & = & \varPhi([H_{\widehat{i}1}\ H_{\widehat{i}2}])\cdot X_{0i,(M_{1}+M_{2}-N_{\widehat{i}})\times d_{0i}^{Z}}^{Z}\label{eq:vvv4}\\
V_{0i}^{R} & = & X_{0i,(M_{1}+M_{2})\times d_{0i}^{R}}^{R}\label{eq:vvv5}\\
V_{i}^{R} & = & X_{i,M_{i}\times d_{i}^{R}}^{R}\label{eq:vvv6}\\
V_{0}^{R} & = & X_{0,(M_{1}+M_{2})\times d_{i}^{R}}^{R}\label{eq:vvv7}
\end{eqnarray}
where $i,j=1,2$ and $\widehat{i}=3-i.$ The beamformers used for
$V_{ij}^{Z}$, $V_{ij}^{A}$ and $V_{0i}^{Z}$ come from the nullspace
of the channels or concatenated channels, and the beamformers used
for $V_{ij}^{R}$, $V_{0i}^{R}$, $V_{i}^{R}$ and $V_{0}^{R}$ are
just generated randomly as described previously. It's shown later
in Section \ref{sec:Achievability-of-the} that, if the DoF tuple
$\boldsymbol{\overrightarrow{d}}$ is in the region of $\mathbb{D}_{\mathrm{in}}$,
then using the above precoding beamformers, all messages are decodable
at their intended receivers with probability one. Hence, the DoF tuple
$\boldsymbol{\overrightarrow{d}}$ is achievable and $\mathbb{D}_{\mathrm{in}}$
is an achievable DoF region. $\ \ \ \ \ \ \ \ \ \ \ \ $ $\blacksquare$
\begin{rem}
In linear beamforming, to achieve $\mathbb{D}_{\mathrm{in}}$ , only
the techniques of zero-forcing, interference alignment and random
beamforming are required. Furthermore, as is shown later in Section
\ref{sec:Achievability-of-the}, interference alignment is needed
only among the four private $X$ channel messages, i.e., aligning $W_{11}$ with $W_{12}$
at receiver $R_{2}$ or aligning $W_{21}$ with $W_{22}$ at receiver
$R_{1}$. Somewhat surprisingly perhaps, it is not necessary to align
interference due to any part of $W_{01}$ with that due to $W_{11}$
or $W_{12}$ at receiver $R_{2}$, or to align interference due to
any part of $W_{02}$ with that due to $W_{21}$ or $W_{22}$ at receiver
$R_{1}$.
\end{rem}

\begin{rem}
\label{rem:Random_combinations} In the construction of $V_{ij}^{Z}$,
$V_{ij}^{A}$ and $V_{0i}^{Z}$, we use the random linear combinations
of the basis vectors of the nullspace of corresponding channels, instead
of directly picking beamformers from those basis vectors obtained
through an SVD. The advantage is that it avoids picking a same basis
vector repetitively in following procedures and potentially leading
to unexpected dependence among the beamformers. 
\end{rem}
\begin{figure*}
\begin{centering}
\includegraphics[width=15cm]{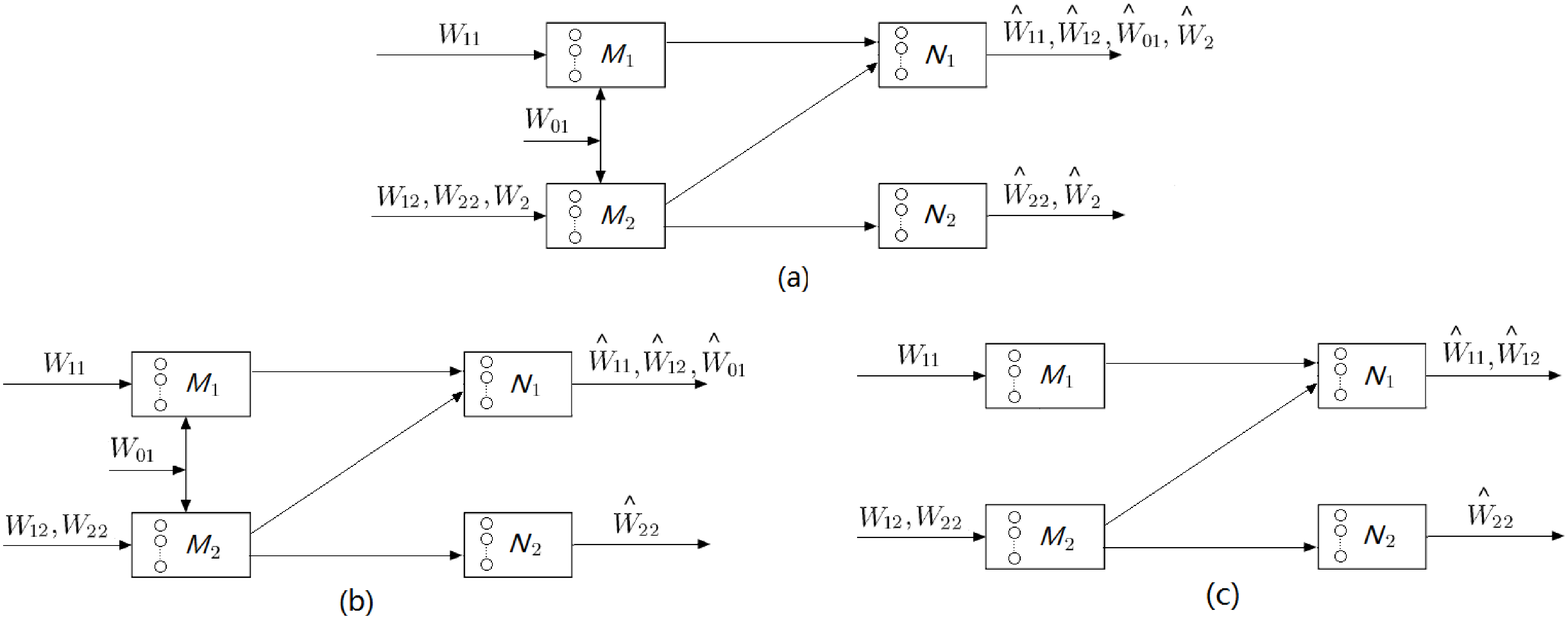}
\par\end{centering}

\caption{\label{fig:MIMO-Z-channel}MIMO $Z_{21}^{*}$ channel with general
message sets (a) complete (b) reduced (c) only private}
\end{figure*}

\begin{lem}
\label{lem:The-fractional-numbers} The fractional numbers at the
boundary of $\mathbb{D}$, i.e., the gap between $\mathbb{D}_{\mathrm{in}}$
and $\mathbb{D}$, can be achieved using appropriate length of symbol
extension. In the case that $M_{1}+M_{2}=N_{1}+N_{2}$ and $\min(M_{1},M_{2},N_{1},N_{2})=1$,
ACS is required in addition to symbol extension.\end{lem}
\begin{IEEEproof}
To achieve a DoF tuple $\boldsymbol{\overrightarrow{d}}$ with fractional
values, we use a $T$ symbol extensions of the channel such that $T\cdot\boldsymbol{\overrightarrow{d}}$
is integer-valued. The problem of unexpected dependencies, which is brought 
on by the structured channel matrices after symbol extensions, also exists here in the 
nine-message problem. The random beamforming part of messages in Groups 2 and 3,
i.e., $W_{0i}^{R}$ and $W_{k}^{R}$ ($i=1,2$, $k=0,1,2$), 
cause no problem; they behave the same as do $W_{ij}^{R}$($i,j=1,2$)
from Group 1 in terms of independence results. Since the beamformers
for the zero-forcing part of Group 2 messages, i.e., $W_{0i}^{Z}$
($i=1,2)$, are generated from the null space of corresponding concatenated
channels, they face the same situation as the interference alignment
beamformer pairs (of the private messages) do. Since all
the zero-forcing and interference alignment beamformers are derived
from the same source but belong to different messages, their behaviors
are actually equivalent when considering independence results. The
analyses of when ACS is necessary and how ACS works which were detailed 
in Section \ref{sec:The-MIMO-X} for the the MIMO $X$ channel are the same 
as in the MIMO $X$ channel problem as well.
\end{IEEEproof}
In summary, Lemmas \ref{lem:An-inner-bound} and \ref{lem:The-fractional-numbers}
establish that $\mathbb{D}$ is an inner bound to the DoF region of
the $2 \times 2$ interference network. Together with the proof of
the outer bound in Section \ref{sec:Outerbound-on-the}, this completes
the proof of Theorem \ref{thm:The-total-degrees}.

\section{\label{sec:Outerbound-on-the}Outerbound on the degrees of freedom
region}

In this section, we prove the converse part of Theorem \ref{thm:The-total-degrees},
i.e., that the region $\mathbb{D}$ is an outer bound for the DoF
region of the $2 \times 2$ interference network.

First, the outer bound (\ref{eq:out9}) comes from the MIMO point-to-point
channel outer bound when cooperation between transmitters and receivers
are both allowed. 

Second, consider the embedded multiple-access channel which only contains
transmitters $T_{1}$ and $T_{2}$ and receiver $R_{1}$. In this
situation, message $W_{02}$, $W_{21}$, $W_{22}$ are irrelevant
and set to $\emptyset$ to avoid interference. The original message
$W_{1}$ will degenerate to $W_{11}$, since we don't require $W_{1}$ 
to be decoded by receiver
$R_{2}$. Similarly, $W_{2}$ will degenerate to $W_{12}$, and $W_{0}$
will degenerate to $W_{01}$. The cut-set bound for multiple-access channel
with common message is $\hat{d}_{01}+\hat{d}_{11}+\hat{d}_{12}\leq N_{1}$.
Hence in this scenario, we get the equivalent outer bound $(d_{0}+d_{01})+(d_{1}+d_{11})+(d_{2}+d_{12})\leq N_{1}$,
which is outer bound (\ref{eq:out5}). In the same way, we get outer
bound (\ref{eq:out6}) by considering the embedded multiple-access
channel which only contains transmitter $T_{1}$ and $T_{2}$ and
receiver $R_{2}$.

Third, consider the embedded broadcast channel which only contains
transmitter $T_{1}$ and receivers $R_{1}$ and $R_{2}$. In this
situation, message $W_{12}$, $W_{22}$ and $W_{2}$ are irrelevant
and set to $\emptyset$ to avoid interference. We also set $W_{0}$,
$W_{01}$, $W_{02}$ to $\emptyset$ and loosen the requirement for
transmitter $T_{1}$ by not requiring it to help in transmitting $W_{0}$,
$W_{01}$ and $W_{02}$. Then we get the outer bound from the result
of broadcast channel with common message $d_{1}+d_{11}+d_{21}\leq M_{1}$,
which is outer bound (\ref{eq:out7}). Similarly, by considering the
embedded broadcast channel with transmitter $T_{2}$, we get outer
bound (\ref{eq:out8}).

Next we prove outer bound (\ref{eq:out2}). Outer bounds (\ref{eq:out1}),
(\ref{eq:out3}) and (\ref{eq:out4}) can be similarly inferred. Consider
the channel depicted in Figure \ref{fig:MIMO-Z-channel}.(a), in which
there is no communication link between transmitter $T_{1}$ and receiver
$R_{2}$. Since channel is the MIMO $2 \times 2$ interference
network with channel matrix $H_{21}=0$, we refer to it as the MIMO
$Z_{21}$ channel. The reduced message sets shown in Figure \ref{fig:MIMO-Z-channel}.(a)
contains all five possible messages for this channel. Thus, Figure
\ref{fig:MIMO-Z-channel}.(a) depicts the $Z_{21}$ channel with fully
general message sets. Here we use {*} to indicate considering fully
general message sets and rename Figure \ref{fig:MIMO-Z-channel}.(a)
as $Z_{21}^{*}$ channel.

We show that the outer bound on the total DoF of the MIMO $Z_{21}^{*}$
channel is also an outer bound of the sum-DoF in the outer bound (\ref{eq:out2})
for the original MIMO $2 \times 2$ interference network with general
message sets, i.e.,
\begin{align}
\underset{D^{2\times2}}{\max}(d_{0}+d_{01}+d_{1}+d_{11}+d_{12}+d_{2}+d_{22})\nonumber \\
\leq\underset{D^{Z_{21}^{*}}}{\max}(d_{01}+d_{11}+d_{12}+d_{2}+d_{22}).\label{eq:2by2Z}
\end{align}

Suppose we have a coding scheme that is able to achieve ($d_{0}$,
$d_{01}$, $d_{1}$, $d_{11}$, $d_{12}$, $d_{2}$, $d_{22}$) on
the nine-message MIMO $2 \times 2$ interference network.
Now, suppose, in place of message $W_{21}$ and $W_{02}$ we use two
known sequences that are available to all transmitters and receivers
\textit{a priori}. Also, a genie provides $W_{11}$, $W_{1}$, $W_{0}$
and $W_{01}$ to receiver $R_{2}$. Thus receiver $R_{2}$ knows all
the information available to transmitter $T_{1}$ and can subtract
transmitter $T_{1}$'s signal from its received signal. This is equivalent
to $H_{21}=\mathbf{0}$. Since receiver $R_{2}$ already knows $W_{1}$,
transmitter $T_{1}$ only needs to make sure that receiver $R_{1}$
can successfully decode $W_{1}$, so that $W_{1}$ degenerates to $W_{11}$.
Similarly, $W_{0}$ degenerates to $W_{01}$. The resulting $2 \times 2$
interference network becomes identical to
the $Z^{*}$ channel with the general message set as depicted in Figure
\ref{fig:MIMO-Z-channel}.(a). Since neither setting $W_{21}$ and
$W_{02}$ to known sequences nor the assistance of genie to receiver
$R_{2}$ can deteriorate the performance of the coding scheme, the
same degrees of freedom $d_{01,Z_{21}^{*}}=d_{01}+d_{0}$, $d_{11,Z_{21}^{*}}=d_{1}+d_{11}$,
$d_{12,Z_{21}^{*}}=d_{21}$, $d_{2,Z_{21}^{*}}=d_{2}$, $d_{22,Z_{21}^{*}}=d_{22}$
are achievable on the $Z^{*}$ channel as well. This proves inequality
(\ref{eq:2by2Z}). The argument here is similar to the proof of Lemma
1 in \cite{jafar2008degrees}, in which $Z_{21}$ channel with only
private messages is considered.

In the $Z_{21}^{*}$ channel depicted in Figure \ref{fig:MIMO-Z-channel}.(a),
message $W_{2}$ is sent out from transmitter 2 and desired at both
receivers, $R_{1}$ and $R_{2}$. If we loosen this requirement
and only demand receiver $R_{2}$  to be able to decode this message,
the degrees of freedom of the new system will be no less than that
of the original system, since reducing decoding requirement cannot
hurt. In this case, $W_{2}$ actually plays the same role as $W_{22}$
does. As a result, we can combine them together and the system reduces
to Figure \ref{fig:MIMO-Z-channel}.(b). 

The system in Figure \ref{fig:MIMO-Z-channel}.(c) is the ordinary
MIMO $Z$ channel, which only contains private messages $W_{11}$,
$W_{12}$, $W_{22}$. An outer bound of ordinary MIMO $Z$ channel
is given in Corollary 1 of \cite{jafar2008degrees}, which is
\begin{gather*}
\max(d_{11}+d_{12}+d_{22})\leq\max(N_{1},M_{2}).
\end{gather*}
The idea of the proof therein is to show the {\em sum} capacity of $Z_{21}$
channel (Figure \ref{fig:MIMO-Z-channel}.(c)) is bounded above by
the MAC with $M_{2}$ receive antennas if $N_{1}<M_{2}$ and bounded
above by the MAC with $N_{1}$ receive antennas if $N_{1}\geq M_{2}$.
The multiplexing gain of a MAC cannot be greater than the total number
of receive antennas. Therefore, we have $\max(d_{11}+d_{12}+d_{22})\leq\max(N_{1},M_{2})$
for Figure \ref{fig:MIMO-Z-channel}.(c). Now, consider the $Z_{21}$
channel in Figure \ref{fig:MIMO-Z-channel}.(b), in which one additional
common message $W_{01}$ is applied. Following the exact same argument
as in \cite{jafar2008degrees} , we get that the {\em sum} capacity
of $Z_{21}$ channel (Figure \ref{fig:MIMO-Z-channel}.(b)) is bounded
above by corresponding MAC with common message, whose multiplexing
gain is also no greater than its total number of receive antennas,
i.e.,
\begin{gather*}
\max(d_{01}+d_{11}+d_{12}+d_{22})\leq\max(N_{1},M_{2}).
\end{gather*}
Including common message or not doesn\textquoteright t affect the
relationship and transformation between $Z$ channel and corresponding
MAC channel in the proof. The reader can refer to \cite{jafar2008degrees}
for more details.

So far we obtained an outer bound for the MIMO $Z^{*}$ channel with general
message sets in Figure \ref{fig:MIMO-Z-channel}.(a), which is
\begin{gather*}
\max(d_{01}+d_{11}+d_{12}+(d_{2}+d_{22}))\leq\max(N_{1},M_{2}).
\end{gather*}
According to inequality (\ref{eq:2by2Z}), we have that an outer bound
for the MIMO $2 \times 2$ interference network with general message
sets is
\begin{gather*}
d_{0}+d_{01}+d_{1}+d_{11}+d_{12}+d_{2}+d_{22}\leq\max(N_{1},M_{2}),
\end{gather*}
which is the outer bound (\ref{eq:out2}).

Similarly, we obtain outer bounds (\ref{eq:out1}), (\ref{eq:out3})
and (\ref{eq:out4}) from the MIMO $Z_{22}^{*}$, $Z_{12}^{*}$, $Z_{11}^{*}$
channel respectively. The general message set for the MIMO $Z_{ij}^{*}$
($i,j\in\{1,2\}$) channel consists of message $W_{\widehat{i}j}$,
$W_{i\widehat{j}}$, $W_{\widehat{i}\widehat{j}}$, $W_{0\widehat{i}}$
and $W_{\widehat{j}}$, where $\widehat{i}=3-i$, $\widehat{j}=3-j$.

\section{\label{sec:Achievability-of-the}Achievability of the inner bound}

We have already described the precoding scheme and given the expressions
for all the beamformers for all nine messages in the outline of proof
of Lemma \ref{lem:An-inner-bound}. In this section, we continue the
proof and show that, using this scheme, the inner bound $\mathbb{D}_{\mathrm{in}}=\textrm{co}\left(\mathbb{D}\cap\mathbb{Z}_{+}^{9}\right)$
is achievable. 

First, it is shown that all the desired messages are distinguishable,
at their intended receivers; and then, we show
that the region achievable is identical to $\mathbb{\mathbb{D}_{\mathrm{in}}}$.

\subsection{Independence requirements}

The messages received by each receiver can be divided into two groups
based on whether they are desired or undesired messages. The undesired
messages are also potentially sources of interference. For receiver
$R_{1}$, desired messages contain $W_{D1}$$=$$(W_{11}$, $W_{12}$,
$W_{01}$, $W_{1}$, $W_{2}$, $W_{0})$, and undesired messages contain
$W_{U1}$$=$$(W_{21}$, $W_{22}$, $W_{02})$. For receiver $R_{2}$,
desired messages contain $W_{D2}$$=$$(W_{21}$, $W_{22}$, $W_{02}$,
$W_{1}$, $W_{2}$, $W_{0})$, and undesired messages contain $W_{U2}$$=$$(W_{11}$,
$W_{12}$, $W_{01})$. Let $D_{i}$ denote the matrix of received
vectors associated with the desired messages at receiver $i$, and
$U_{i}$ denote the matrix of directions of the receive beamformers
associated with the undesired messages at receiver $i$. We thus have
\begin{align*}
D_{1} & =\biggl[H_{11}V_{11}\,|\,H_{12}V_{12}\,|\,\left[H_{11}\ H_{12}\right]V_{01}\,|\,\cdots\\
 & \ \ \ \ \ \ \ \ \ \cdots\ H_{11}V_{1}\,|\,H_{12}V_{2}\,|\,\left[H_{11}\ H_{12}\right]V_{0}\biggr]\\
D_{2} & =\biggl[H_{21}V_{21}\,|\,H_{22}V_{22}\,|\,\left[H_{21}\ H_{22}\right]V_{02}\,|\,\cdots\\
 & \ \ \ \ \ \ \ \ \ \cdots\ H_{21}V_{1}\,|\,H_{22}V_{2}\,|\,\left[H_{21}\ H_{22}\right]V_{0}\biggr]\\
U_{1} & =\biggl[H_{11}V_{21}\,|\,H_{12}V_{22}\,|\,\left[H_{11}\ H_{12}\right]V_{02}\biggr]\\
U_{2} & =\biggl[H_{21}V_{11}\,|\,H_{22}V_{12}\,|\,\left[H_{21}\ H_{22}\right]V_{01}\biggr].
\end{align*}
For successful communication, each receiver needs to be able to decode
all its own desired messages. In order to take the most advantage
of channel resource, we allocate as much resource as possible to desired
messages to minimize the resource consumed by undesired messages,
i.e., by interference. 
\begin{lem}
\label{lem:full_rank_condition}If\textup{ }all the channels are generic\textup{
}and the following constraints are satisfied\textup{
\begin{align}
 & d_{1}+d_{2}+d_{0}+d_{01}+d_{11}+d_{12}+d_{21}+d_{22}+d_{02}\nonumber \\
 & \ \ \ \ \ \ \ \ \ \ \ \ \ \ \ -Z_{21}-Z_{22}-A_{1}-Z_{02}\leq N_{1}\label{eq:constraint1}\\
 & d_{1}+d_{2}+d_{0}+d_{02}+d_{21}+d_{22}+d_{11}+d_{12}+d_{01}\nonumber \\
 & \ \ \ \ \ \ \ \ \ \ \ \ \ \ \ -Z_{11}-Z_{12}-A_{2}-Z_{01}\leq N_{2}\label{eq:constraint2}\\
 & d_{1}+d_{11}+d_{21}\leq M_{1}\label{eq:constraint3}\\
 & d_{2}+d_{12}+d_{22}\leq M_{2}\label{eq:constraint4}\\
 & d_{1}+d_{2}+d_{0}+d_{01}+d_{02}+d_{11}+d_{21}+d_{12}+d_{22}\nonumber \\
 & \ \ \ \ \ \ \ \ \ \ \ \ \ \ \ \leq\min(M_{1}+M_{2},N_{1}+N_{2}),\label{eq:constraint5}
\end{align}
}using the precoding scheme described in the outline of proof of Lemma
\ref{lem:An-inner-bound} in Section \ref{sec:Main-Result}, we have
the following independence results
\begin{flalign}
 & \mathrm{rank}(U_{1})=(d_{21}-Z_{21})+(d_{22}-Z_{22})-A_{1}+(d_{02}-Z_{02})\label{eq:rank_proof_1}\\
 & \mathrm{rank}(U_{2})=(d_{11}-Z_{11})+(d_{12}-Z_{12})-A_{2}+(d_{01}-Z_{01})\label{eq:rank_proof_2}\\
 & \mathrm{rank}(D_{1})=d_{11}+d_{12}+d_{01}+d_{1}+d_{2}+d_{0}\label{eq:rank_proof_3}\\
 & \mathrm{rank}(D_{2})=d_{21}+d_{22}+d_{02}+d_{1}+d_{2}+d_{0},\label{eq:rank_proof_4}\\
 & \mathrm{rank}([D_{1}\ U_{1}])=\mathrm{rank}(D_{1})+\mathrm{rank}(U_{1})\label{eq:rank_proof_5}\\
 & \mathrm{rank}([D_{2}\ U_{2}])=\mathrm{rank}(D_{2})+\mathrm{rank}(U_{2}).\label{eq:rank_proof_6}
\end{flalign}
These independence results together ensure that all desired messages
are distinguishable, and thus decodable, at their intended receivers.
\end{lem}
Note that, constraints (\ref{eq:constraint1}) and (\ref{eq:constraint2})
imply that the total independent number of received beamformers at
receiver $R_{i}$ will be no greater than $N_{i}$, the number of
its antennas; constraints (\ref{eq:constraint3}) and (\ref{eq:constraint4})
imply that the number of independent streams sent out by transmitter
$T_{i}$ are restricted to be no greater than $M_{i}$; constraint
(\ref{eq:constraint5}) implies that the number of all independent
streams two transmitters sent out together will be no greater than
their total number of antennas. 

Regarding the independence result, equations (\ref{eq:rank_proof_1})
and (\ref{eq:rank_proof_2}) give the dimension of the subspace spanned
by the received beamformers associated with the undesired messages,
i.e., interference; equations (\ref{eq:rank_proof_3}) and (\ref{eq:rank_proof_4})
show that the directions of the received beamformers associated with
the desired messages at each receiver are linearly independent of
each other; equations (\ref{eq:rank_proof_5}) and (\ref{eq:rank_proof_6})
indicate that the subspace occupied by the desired messages is linearly
independent of that of the interference. 
\begin{IEEEproof}
We only give the proof of (\ref{eq:rank_proof_1}), (\ref{eq:rank_proof_4})
and (\ref{eq:rank_proof_5}), since the other three follow in the same way.

First consider equation (\ref{eq:rank_proof_1}). According to the
expressions of beamformers provided in equations (\ref{eq:vvv1}),
(\ref{eq:vvv3}) and (\ref{eq:vvv5}), we have that $V_{21}^{Z}$
and $V_{22}^{Z}$ are drawn from the nullspace of $H_{11}$ and $H_{22}$,
respectively, and $V_{02}^{Z}$ is generated from the nullspace $\mathcal{N}([H_{11}\ H_{12}])$.
Thus, they will all be zero-forced at receiver $R_{1}$, i.e.,
\begin{gather*}
H_{11}V_{21}^{Z}=0\\
H_{12}V_{22}^{Z}=0\\{}
[H_{11}\ H_{12}]V_{02}^{Z}=0.
\end{gather*}
Consequently, we have
\begin{eqnarray*}
\mathrm{rank}(H_{11}V_{21}) & = & \mathrm{rank}(H_{11}[V_{21}^{Z}\ V_{21}^{A}\ V_{21}^{R}])\\
 & = & \mathrm{rank}(H_{11}[V_{21}^{A}\ V_{21}^{R}])\\
\mathrm{rank}(H_{12}V_{22}) & = & \mathrm{rank}(H_{12}[V_{22}^{Z}\ V_{22}^{A}\ V_{22}^{R}])\\
 & = & \mathrm{rank}(H_{12}[V_{22}^{A}\ V_{22}^{R}])\\
\mathrm{rank}(\left[H_{11}\ H_{12}\right]V_{02}) & = & \mathrm{rank}(\left[H_{11}\ H_{12}\right][V_{02}^{Z}\ V_{02}^{R}])\\
 & = & \mathrm{rank}(\left[H_{11}\ H_{12}\right]V_{02}^{R}).
\end{eqnarray*}
Furthermore, from equation (\ref{eq:vvv2}), we have
\begin{eqnarray*}
 &  & [H_{11}\ H_{12}]\left[\begin{array}{c}
V_{21}^{A}\\
V_{22}^{A}
\end{array}\right]=0\\
\Rightarrow &  & \ \ H_{11}V_{21}^{A}+H_{12}V_{22}^{A}=0,
\end{eqnarray*}
which indicates that the subspace spanned by $H_{11}V_{21}^{A}$ is
aligned with the subspace spanned by $H_{12}V_{22}^{A}$ at receiver
$R_{1}$. So, we have
\begin{eqnarray*}
\mathrm{rank}([H_{11}V_{21}^{A}\ H_{12}V_{22}^{A}]) & = & \mathrm{rank}(H_{11}V_{21}^{A})\\
 & = & \mathrm{rank}(H_{12}V_{22}^{A}).
\end{eqnarray*}
One can observe that the nullspace of $H_{11}$ and $H_{12}$ is closely
related to the nullspace of $[H_{11}\ H_{12}]$. In particular, since
$H_{11}\varPhi(H_{11})=0$ and $H_{12}\varPhi(H_{12})=0$, we have
that
\begin{eqnarray*}
[H_{11}\ H_{12}]\left[\begin{array}{c}
\varPhi(H_{11})\\
0
\end{array}\right] & = & 0\\{}
[H_{11}\ H_{12}]\left[\begin{array}{c}
0\\
\varPhi(H_{12})
\end{array}\right] & = & 0,
\end{eqnarray*}
which means that the column vectors of $\left[\begin{array}{c}
\varPhi(H_{11})\\
0
\end{array}\right]$ and $\left[\begin{array}{c}
0\\
\varPhi(H_{12})
\end{array}\right]$ are both in $\mathcal{N}([H_{11}\ H_{12}])$. Since beamformer $\left[\begin{array}{c}
V_{21}^{A}\\
V_{22}^{A}
\end{array}\right]$ is obtained as random linear combinations of the null space basis
vectors $\varPhi([H_{11}\ H_{12}])$, the probability that it belongs
to the subspace spanned only by column vectors of $\left[\begin{array}{c}
\varPhi(H_{11})\\
0
\end{array}\right]$ and $\left[\begin{array}{c}
0\\
\varPhi(H_{12})
\end{array}\right]$ is zero. In other words, $[H_{11}V_{21}^{A}]$ and $[H_{12}V_{22}^{A}]$
will have full column rank almost surely, since none of the column vectors
of $V_{21}^{A}$ or $V_{22}^{A}$ will be accidentally zero-forced
at receiver $R_{1}$. This is one benefit of using random linear combinations,
as mentioned in Remark \ref{rem:Random_combinations}. 

Beamformers $V_{21}^{R}$, $V_{22}^{R}$ and $V_{02}^{R}$ are generated
randomly, they will all have full column rank almost surely. Their
projections at the receivers will be linearly independent of each
other unless they can't be. According to constraint (\ref{eq:constraint1})-(\ref{eq:constraint5}),
the total number of beamformers transmitted in any channel is always
no greater than the channel dimension, so there will be no loss of
column ranks. As a result, we have
\begin{flalign*}
 & \mathrm{rank}(U_{1})=\mathrm{rank}(\left[H_{11}V_{21}^{A}\ H_{11}V_{21}^{R}\ H_{12}V_{22}^{R}\ [H_{11}\ H_{12}]V_{02}^{R}\right])\\
 & =A_{1}+(d_{21}-Z_{21}-A_{1})+(d_{22}-Z_{22}-A_{1})+(d_{02}-Z_{02}),
\end{flalign*}
which proves equation (\ref{eq:rank_proof_1}). Similarly, we have
equation (\ref{eq:rank_proof_2}).

Next, consider equation (\ref{eq:rank_proof_4}). We have just shown
that sending a symbol of $W_{21}^{Z}$ or $W_{22}^{Z}$ or $W_{02}^{Z}$,
or a pair of symbols of $W_{21}^{A}$ and $W_{22}^{A}$ will consume
1 dimension of the subspace of $[H_{11}\ H_{12}]$. From the dimension
of each part given in equations (\ref{eq:ddd1}), (\ref{eq:ddd2})
and (\ref{eq:ddd4}), we have that
\begin{flalign}
Z_{21} & \leq(M_{1}-N_{1})^{+}\label{eq:nullspace1}\\
Z_{22} & \leq(M_{2}-N_{1})^{+}\label{eq:nullspace2}\\
Z_{21}+Z_{22}+A_{1}+Z_{02} & \leq(M_{1}+M_{2}-N_{1})^{+},\label{eq:nullspace3}
\end{flalign}
which means the total numbers of beamformers do not exceed the dimensions
of corresponding nullspaces. Since we generate all the beamformers
as random linear combinations of the entire basis of the respective
nullspaces, column vectors of $V_{A}=\left[\begin{array}{c}
V_{21}^{Z}\\
0
\end{array}\Biggl|\begin{array}{c}
0\\
V_{22}^{Z}
\end{array}\Biggl|\,\begin{array}{c}
V_{21}^{A}\\
V_{22}^{A}
\end{array}\,\Biggl|\,V_{02}^{Z}\right]$ will be linearly independent of each other almost surely. Meanwhile,
they will also be linearly independent of the random column vectors
of $V_{B}=\left[\begin{array}{c}
V_{21}^{R}\\
0
\end{array}\Biggl|\begin{array}{c}
0\\
V_{22}^{R}
\end{array}\,\Biggl|\,V_{02}^{R}\,\Biggl|\,V_{1}^{R}\,\Biggl|\,V_{2}^{R}\,\Biggl|\,V_{0}^{R}\right]$. Since all of these beamformers in $V_{A}$ and $V_{B}$ are derived
from $[H_{11}\ H_{12}]$ or generated randomly, they are independent
of channel matrix $[H_{21}\ H_{22}]$. Since $H_{21}$ and $H_{22}$
are both full rank matrices with generic elements, the column vectors
of $[H_{21}\ H_{22}]\left[V_{A}\ V_{B}\right]$ will be linearly dependent
only if they have to be linearly dependent. Because we have the constraint
(\ref{eq:constraint2}), which indicates $d_{21}+d_{22}+d_{02}+d_{1}+d_{2}+d_{0}\leq N_{2}$,
$[H_{21}\ H_{22}]\left[V_{A}\ V_{B}\right]$ will have rank $d_{21}+d_{22}+d_{02}+d_{1}+d_{2}+d_{0}$
almost surely. So, we have equation (\ref{eq:rank_proof_4}). Similarly,
we have equation (\ref{eq:rank_proof_3}).

Finally, consider equation (\ref{eq:rank_proof_5}). Since the beamformers
associated with $D_{1}$ are independent of the beamformers associated
with $U_{1}$, the subspace spanned by $D_{1}$ and the subspace spanned
by $U_{1}$ will be linearly dependent only if they have to be linearly
dependent. According to constraint (\ref{eq:constraint1}), $\mathrm{rank}(D_{1})+\mathrm{rank}(U_{1})\leq N_{1}.$
Consequently , $\mathrm{rank}([D_{1}\ U_{1}])$ will be equal to $\mathrm{rank}(D_{1})+\mathrm{rank}(U_{1})$
almost surely. So we have equation (\ref{eq:rank_proof_5}). Similarly,
we have equation (\ref{eq:rank_proof_6}).
\end{IEEEproof}
In Lemma \ref{lem:full_rank_condition}, we show that if inequalities
(\ref{eq:constraint1})-(\ref{eq:constraint5}) are satisfied, all
desired messages will be distinguishable at their respectively intended
receivers. In other words, DoF tuples that satisfy (\ref{eq:constraint1})-(\ref{eq:constraint5})
are achievable. In the next section, we explicitly characterize this
achievable DoF region.

\subsection{The achievability of inner bound}

According to the analysis in Lemma \ref{lem:full_rank_condition}
of the precoding scheme described in Section \ref{sec:Main-Result},
we have shown the achievability of the integer-valued points in $\mathbb{D}_{\mathrm{eq}}$,
which is defined as
\begin{align}
\mathbb{D}_{\mathrm{eq}}\triangleq\Bigl\{ & (d_{11},d_{21},d_{12},d_{22},d_{1},d_{2},d_{01},d_{02},d_{0})\in\mathbb{R}_{+}^{\mathsf{E}}:\nonumber \\
 & d_{1}+d_{2}+d_{0}+d_{01}+d_{11}+d_{12}+d_{21}+d_{22}+d_{02}\nonumber \\
 & \quad\quad\quad\quad\quad-Z_{21}-Z_{22}-A_{1}-Z_{02}\leq N_{1}\label{eq:achievable1}\\
 & d_{1}+d_{2}+d_{0}+d_{02}+d_{21}+d_{22}+d_{11}+d_{12}+d_{01}\nonumber \\
 & \quad\quad\quad\quad\quad-Z_{11}-Z_{12}-A_{2}-Z_{01}\leq N_{2}\label{eq:achievable2}\\
 & d_{1}+d_{11}+d_{21}\leq M_{1}\label{eq:achievable7}\\
 & d_{2}+d_{12}+d_{22}\leq M_{2}\label{eq:achievable8}\\
 & d_{1}+d_{2}+d_{0}+d_{01}+d_{02}+d_{11}+d_{21}+d_{12}+d_{22}\nonumber \\
 & \quad\quad\quad\quad\quad\quad\leq\min(M_{1}+M_{2},N_{1}+N_{2})\label{eq:achievable9}\\
 & \textrm{are\,\ satisfied\,\ for\,\ some}\nonumber \\
 & \{(Z_{11},Z_{12},Z_{21},Z_{22},A_{1},A_{2},Z_{01},Z_{02})\in\mathbb{R}_{+}^{\mathsf{A}}:\nonumber \\
 & Z_{21}+Z_{22}+A_{1}+Z_{02}\leq(M_{1}+M_{2}-N_{1})^{+}\label{eq:aux_1}\\
 & Z_{21}\leq(M_{1}-N_{1})^{+}\label{eq:aux_2}\\
 & Z_{22}\leq(M_{2}-N_{1})^{+}\label{eq:aux_3}\\
 & Z_{21}+A_{1}\leq d_{21}\label{eq:aux_4}\\
 & Z_{22}+A_{1}\leq d_{22}\label{eq:aux_5}\\
 & Z_{02}\leq d_{02}\label{eq:aux_6}\\
 & Z_{11}+Z_{12}+A_{2}+Z_{01}\leq(M_{1}+M_{2}-N_{2})^{+}\label{eq:aux_7}\\
 & Z_{11}\leq(M_{1}-N_{2})^{+}\label{eq:aux_8}\\
 & Z_{12}\leq(M_{2}-N_{2})^{+}\label{eq:aux_9}\\
 & Z_{11}+A_{2}\leq d_{11}\label{eq:aux_10}\\
 & Z_{12}+A_{2}\leq d_{12}\label{eq:aux_11}\\
 & Z_{01}\leq d_{01}\}\Bigr\}\label{eq:aux_12}
\end{align}
where set $\mathsf{A}$ contains all the auxiliary variables. Inequalities
(\ref{eq:aux_1})-(\ref{eq:aux_12}) on the auxiliary variables are
obtained from equations (\ref{eq:ddd1})-(\ref{eq:ddd6}). 

To prove the inner bound, we need to find the connection between $\mathbb{D}$
and $\mathbb{D}_{\mathrm{eq}}$. Interestingly, it is shown that these
two regions are identical. However, note that
$\mathbb{D}_{\mathrm{eq}}$ is obtained from a 17-dimensional polyhedron
in $\mathbb{R}_{+}^{\mathsf{E}}\times\mathbb{R}_{+}^{\mathsf{A}}$
defined via 17 inequalities which include eight auxiliary variables.
The problem is to project this polyhedron onto the nine dimensional
positive orthant $\mathbb{R}_{+}^{\mathsf{E}}$. The standard technique
to perform this projection is via the Fourier-Motzkin
Elimination wherein the auxiliary variables are eliminated one at
a time but by creating a large number of inequalities of $\mathrm{O}(m^{2})$
starting with $m$ inequalities and then eliminating redundant inequalities
\cite{ElGamal2011}. Such a technique is clearly infeasible for the
size of the problem at hand here. Instead, we use the special structure
of the inequalities that define $\mathbb{D}_{\mathrm{eq}}$ to prove
that it is equivalent to $\mathbb{D}$ in the following lemma.
\begin{lem}
\label{lem:The-9-dimensional-region}The 9-dimensional region $\mathbb{D}$
is equal to \textup{$\mathbb{D}_{\mathrm{eq}}$.}\end{lem}
\begin{IEEEproof}
First show any vector in $\mathbb{D}$ is also in $\mathbb{D}_{\mathrm{eq}}$,
and then show any vector in $\mathbb{D}_{\mathrm{eq}}$ is also in
$\mathbb{D}$. The detailed proof is given in Appendix \ref{sec:Equivalence-of-DD}. 
\end{IEEEproof}
Thus, we prove that the inner bound $\mathbb{D}_{\mathrm{in}}=\textrm{co}\left(\mathbb{D}\cap\mathbb{Z}_{+}^{9}\right)$
is achievable.
\begin{figure*}
\begin{centering}
\includegraphics[width=14cm]{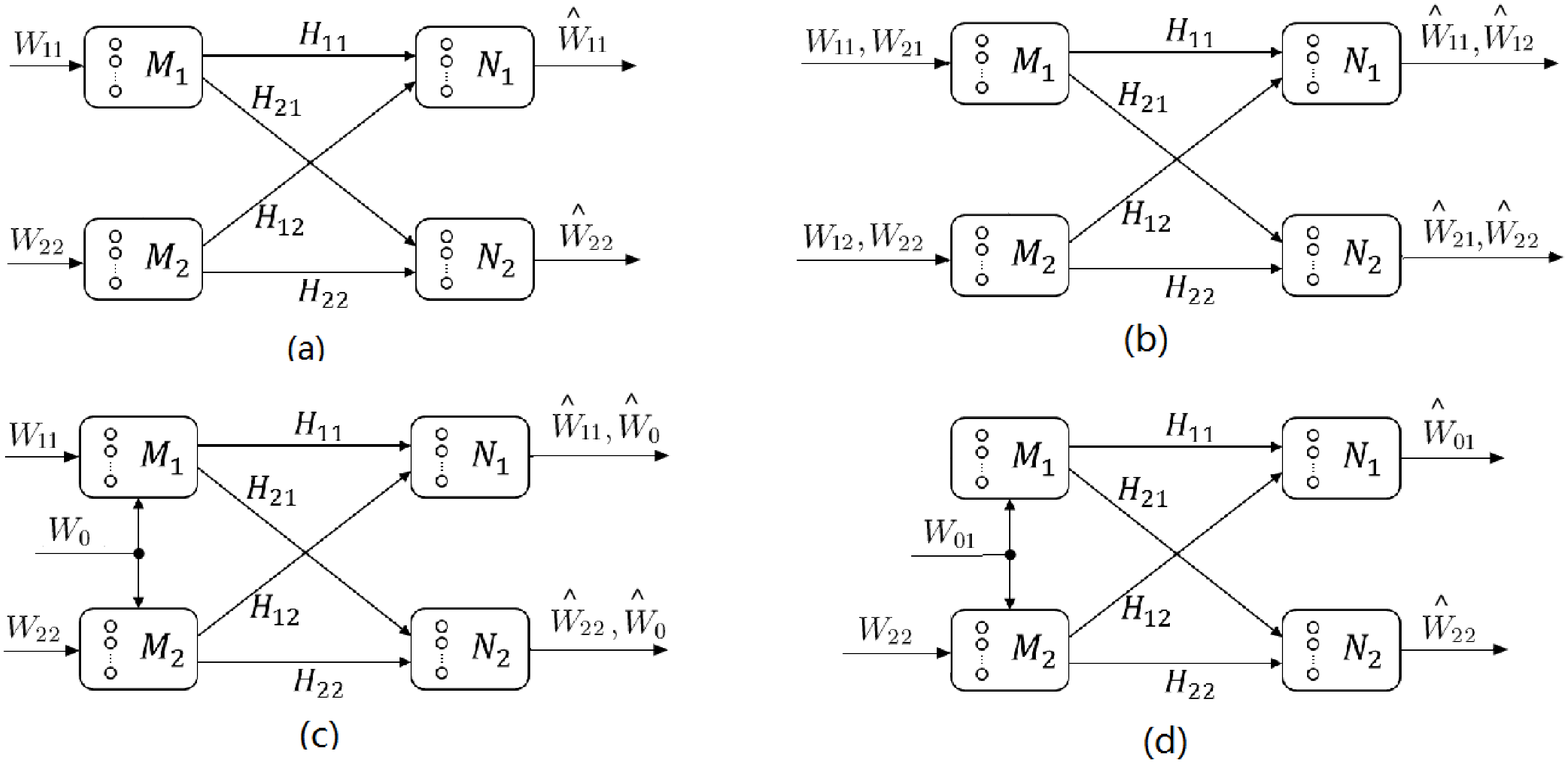}
\par\end{centering}

\caption{\label{fig:MIMO-(a)-IC} several MIMO $2 \times 2$ interference networks
(a) IC (b) IC-CM (c) cognitive IC}
\end{figure*}

\subsection{\label{sub:No-interference-alignment}No interference alignment is
needed for $W_{01}$ and $W_{02}$}

In our precoding scheme, interference alignment is used only among
the four private messages. Only zero-forcing is used for the cognitive and common messages $W_{01}$ and $W_{02}$. In this section, we demonstrate why.

Consider $W_{02}$, for instance. If $M_{1}+M_{2}>N_{1}$, transmit zero-forcing
of $W_{02}$ is possible. We can choose beamformers for $W_{02}$
from the null space $\mathcal{N}([H_{11}\,\,H_{12}])$. It is worth
noting that $\mathcal{N}([H_{11}\,\,H_{12}])$ has already been used
to generate $V_{21}^{Z}$, $V_{22}^{Z}$ and ($V_{21}^{A}$, $V_{22}^{A}$)
pairs. To transmit a data symbol in $W_{02}$, we cannot choose a
vector in the span of the column vectors in $\left[\begin{array}{c}
V_{21}^{Z}\\
0
\end{array}\right]$, $\left[\begin{array}{c}
0\\
V_{22}^{Z}
\end{array}\right]$ and $\left[\begin{array}{c}
V_{21}^{A}\\
V_{22}^{A}
\end{array}\right]$, otherwise the data symbol of $W_{02}^{Z}$ will not be distinguishable
with part of $W_{21}^{Z}$, $W_{22}^{Z}$ and ($W_{21}^{A}$, $W_{22}^{A}$)
at receiver $R_{2}$. As a result, $V_{02}^{Z}$ can be only chosen
from the unoccupied subspace of $\mathcal{N}([H_{11}\,\,H_{12}])$.
This is also why the dimension available for transmit zero-forcing
of $W_{02}^{Z}$ is at most $M_{1}+M_{2}-N_{1}-d_{21}^{Z}-d_{22}^{Z}-A_{1}$
in equation (\ref{eq:ddd4}).

Next, consider the possibility of aligning the beamformer, denoted
as $V_{02}^{A}$, of data symbol in $W_{02}$ with the existing interference
due to $W_{21}^{R}$, $W_{22}^{R}$ or ($W_{21}^{A},$ $W_{22}^{A}$).
Take ($W_{21}^{A},$ $W_{22}^{A}$) for example. If vector $[H_{11}\,\,H_{12}]v_{02}^{A}$
aligns with ($H_{11}v_{21}^{A}$, $H_{12}v_{22}^{A}$), where $v_{21}^{A}$
and $v_{22}^{A}$ are some column vectors lie in $\mathrm{span}(V_{21}^{A})$
and $\mathrm{span}(V_{22}^{A})$, respectively, it is easy to see
that
\begin{align*}
v_{02}^{A} & =\left[\begin{array}{c}
\alpha v_{21}^{A}\\
\mathbf{0}
\end{array}\right]+\left[\begin{array}{c}
\mathbf{0}\\
\beta v_{22}^{A}
\end{array}\right]+\gamma v_{0}
\end{align*}
where $\alpha,\beta,\gamma\in\mathbb{C}^{1}$, $v_{0}$ is a column
vector in the null space $\mathcal{N}([H_{11}\,\,H_{12}])$. To make
$W_{02}^{A}$ distinguishable at receiver $R_{2}$, $v_{0}$ must
be linearly independent of the already used subspace of $\mathcal{N}([H_{11}\,\,H_{12}])$.
Hence, if we transmit a data symbol in $W_{02}^{A}$ by having its
direction lie in the subspace spanned by the directions associated
with data symbols in $W_{21}^{A}$ and $W_{22}^{A}$, we consume one
dimension in $\mathcal{N}([H_{11}\,\,H_{12}])$. A similar result holds
in attempting to align with existing interference $H_{11}V_{21}^{R}$
or $H_{12}V_{22}^{R}$ at receiver $R_{1}$.

In summary, for each $W_{02}$ stream, both transmit zero-forcing
and interference alignment consume one more available dimension of
$\mathcal{N}([H_{11}\,\,H_{12}])$. In other words, either strategy
costs the same in terms of using the remaining subspace (if any) of
$\mathcal{N}([H_{11}\,\,H_{12}])$. As a practical matter, one might
choose transmit zero-forcing since it easier to compute the corresponding
beamformer.

\section{Special cases}

In this section, we specify the DoF regions for small special cases of Theorem \ref{thm:The-total-degrees}.
\begin{figure*}
\begin{centering}
\includegraphics[width=14cm]{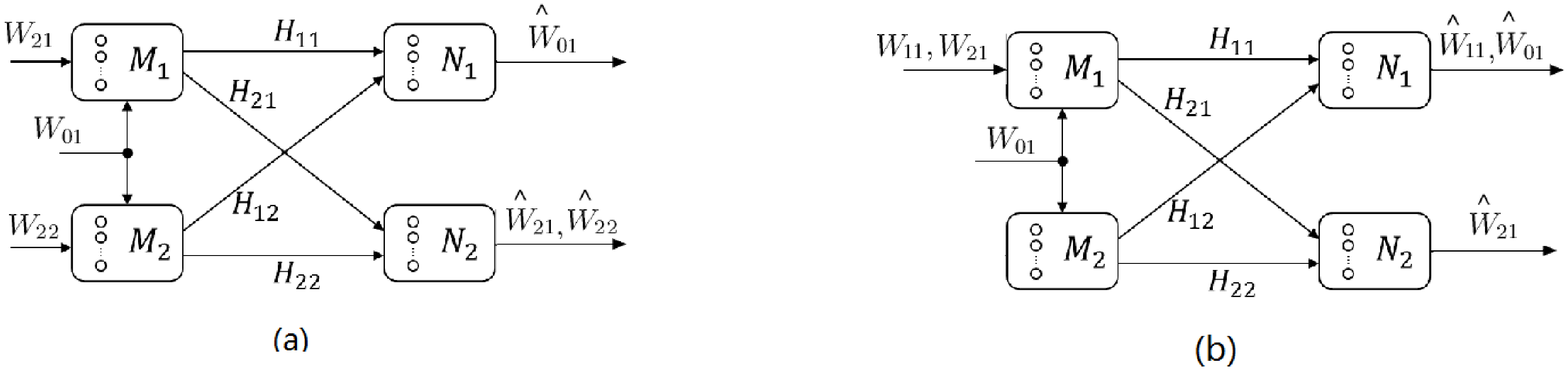}
\par\end{centering}

\caption{\label{fig:special-cases-2}several MIMO $2 \times 2$  interference
networks (a) generalized cognitive IC (b) BC-PCR}
\end{figure*}

\subsection{Known Results as Special Cases}

\subsection*{Case 1. IC (Figure \ref{fig:MIMO-(a)-IC}.a)}

There are only two messages in the interference channel, i.e., $W_{11}$,
$W_{22}$. By eliminating all absent variables in $\mathbb{D}$, we
get the degrees of freedom region for two-user interference channel
as
\begin{align*}
\mathbb{D}_{IC}=\Bigl\{ & (d_{11},d_{22})\in\mathbb{R}_{+}^{2}:\\
 & d_{11}\leq\min(M_{1},N_{1}),\,\,d_{22}\leq\min(M_{2},N_{2}),\\
 & d_{11}+d_{22}\leq\min\left(\max(M_{2},N_{1}),\max(M_{1},N_{2})\right)\Bigr\}
\end{align*}
Hence, Theorem \ref{thm:The-total-degrees} reduces to the well-known
result in \cite{Jafar2007}. We can follow the precoding scheme shown
in Section \ref{sec:Main-Result} and skip the parts that are not
applicable. In this case, we only need to consider $\left[V_{11}^{Z}\,\,V_{11}^{R}\right]$
and $\left[V_{22}^{Z}\,\,V_{22}^{R}\right]$. Only transmit zero-forcing
is possible here. Interference alignment is not applicable since
there is only one source of interference at each receiver.

\subsection*{Case 2. IC-CM (Figure \ref{fig:MIMO-(a)-IC}.b)}

Specializing Theorem \ref{thm:The-total-degrees} to the case where
only messages $W_{11}$, $W_{22}$ and $W_{0}$ are present as depicted
in Fig. \ref{fig:MIMO-(a)-IC}.b (and eliminating absent variables),
we have
\begin{align*}
\mathbb{D}_{IC-CM}=\Bigl\{ & (d_{11},d_{22},d_{01})\in\mathbb{R}_{+}^{3}:\\
 & d_{11}\leq M_{1},\,\,d_{22}\leq M_{2},\\
 & d_{0}+d_{11}\leq N_{1},\,\,d_{0}+d_{22}\leq N_{2},\\
 & d_{0}+d_{11}+d_{22}\leq\\
 & \min\bigl(M_{1}+M_{2},\max(M_{2},N_{1}),\max(M_{1},N_{2})\bigr)\Bigr\}.
\end{align*}
In this case, $W_{11}$ and $W_{22}$ are transmitted using the same
scheme as in IC along with random beamforming for $W_{0}$ 
in the remaining channel dimensions that are still available.

\subsection*{Case 3. Cognitive IC (Figure \ref{fig:MIMO-(a)-IC}.c)}

Theorem \ref{thm:The-total-degrees}, when specialized to the degraded
message set depicted in Fig. \ref{fig:MIMO-(a)-IC}.c, results in
the DoF region of the Cognitive IC, which is
\begin{align*}
\mathbb{D}_{co-IC}=\Bigl\{ & (d_{01},d_{22})\in\mathbb{R}_{+}^{2}:\\
 & d_{01}\leq N_{1},\,\,d_{22}\leq\min(M_{2},N_{2}),\\
 & d_{01}+d_{22}\leq\min\bigl(M_{1}+M_{2},\max(M_{2},N_{1})\bigr)\Bigr\}.
\end{align*}
This DoF region matches with the result of \cite{Huang2009} in the
same cognitive message sharing scenario. In this case, we only need
zero-forcing and random beamforming to achieve any vertex of the DoF region.
The dimensions of symbols of $W_{01}$
and $W_{22}$ that are transmitted using zero-forcing are $\min\left(d_{01},(M_{1}+M_{2}-N_{2})^{+}\right)$
and $\min\left(d_{22},(M_{2}-N_{1})^{+}\right)$, respectively.

\subsection{Examples of New Results }

\subsection*{Case 4. Generalized Cognitive IC (Figure \ref{fig:special-cases-2}.a)}

Consider the generalized cognitive IC, in which there are three messages
$W_{21}$, $W_{01}$ and $W_{22}$. In this model, the two transmitters
send one message each, i.e., $W_{21}$ and $W_{22}$, respectively, 
to Receiver 2 along with another message, i.e., $W_{01}$,
cooperatively to the Receiver 1. Specializing Theorem \ref{thm:The-total-degrees}
to this model, we have the following DoF region result
\begin{align*}
\mathbb{D}_{g-co-IC}=\bigl\{ & (d_{21},d_{22},d_{01})\in\mathbb{R}_{+}^{3}:\\
 & d_{01}\leq N_{1},\,\,d_{21}\leq M_{1},\,\,d_{22}\leq M_{2},\\
 & d_{21}+d_{22}\leq N_{2},\\
 & d_{01}+d_{21}\leq\max(M_{1},N_{1}),\\
 & d_{01}+d_{22}\leq\max(M_{2},N_{1}),\\
 & d_{01}+d_{21}+d_{22}\leq M_{1}+M_{2}\bigr\}.
\end{align*}
Both zero-forcing and interference alignment, if possible, are used
to mitigate the impact of two private messages $W_{21}$ and $W_{22}$
on their common unintended receiver, i.e., receiver $R_{1}$; while
zero-forcing, if possible, is used to reduce the interference received
by receiver $R_{2}$ due to message $W_{01}$.

\subsection*{Case 5. Broadcast Channel with Partially Cognitive Relay (BC-PCR)
(Figure \ref{fig:special-cases-2}.b)}

Consider the model depicted in Figure \ref{fig:special-cases-2}.b.
Transmitter 1 broadcasts two private messages $W_{11}$ and $W_{21}$
to two receivers, respectively, while it simultaneously cooperates
with transmitter 2 (the PCR) to send another message $W_{01}$ to
receiver $R_{1}$. From Theorem \ref{thm:The-total-degrees}, we can
deduce the DoF region of BC-PCR as
\begin{align*}
 & \mathbb{D}_{BC-PCR}=\Bigl\{(d_{21},d_{11},d_{01})\in\mathbb{R}_{+}^{3}:\\
 & \;\;\;\;d_{21}\leq N_{2},\,\,d_{01}+d_{11}\leq N_{1},\,\,d_{11}+d_{21}\leq M_{1},\\
 & \;\;\;\;d_{01}+d_{11}+d_{21}\leq\min\left(M_{1}+M_{2},\max(M_{1},N_{1})\right)\Bigr\}.
\end{align*}
From the analysis in Section \ref{sub:No-interference-alignment}.B.(2),
we know that using zero-forcing, if possible, is enough for transmitting
message $W_{01}$. There is no need to additionally attempt to align
the symbols of $W_{11}$ and $W_{01}$ together at receiver $R_{2}$,
since interference alignment and zero-forcing costs the same in terms
of using the null space of $[H_{21}\,\,H_{22}]$.

\section{Conclusion}

The degrees of freedom region for the nine-message MIMO $2 \times 2$ interference network is established. Each of the nine messages is uniquely identified based on the transmitter(s) it is known to and the receiver(s) at which it is desired and therefore include broadcast/multiple-access/multicast/cognitive/common messages. The DoF region for a setting that involves any subset of the nine messages can thus be derived as a special case. In particular, the DoF region of the MIMO X channel, a problem that remained open despite previous studies, is completely settled.

The achievability scheme uses (a) transmit zero-forcing, a well-known
technique known to be sufficient for the MIMO IC \cite{ElGamal2011},
interference alignment and symbol extensions the necessity (but not
sufficiency) for which was discovered in the context of the constant-coefficient
MIMO $X$ channel in \cite{jafar2008degrees}, and finally, asymmetric
complex signaling which was discovered in the context of the constant-coefficient
SISO $X$ channel in \cite{cadambe2010interference}, but whose benefit
(necessity or sufficiency) in the MIMO (i.e., non-SISO) $X$ channel
remained unclear despite \cite{jafar2008degrees,cadambe2010interference}.
The achievability scheme in this paper combines the principles of
transmit zero-forcing, interference alignment, symbol extensions and
ACS in a novel way that allows not only the complete characterization
of the DoF of the four-message MIMO $X$ channel -- thereby proving
that they are both necessary and sufficient in general for the constant-coefficient
MIMO $X$ channel -- but also the precise DoF region of the much more general
nine-message, constant-coefficient MIMO $2 \times 2$ network considered
in this paper. 

In considering some interesting subsets of the general message set
(including the 9-message case) for the $2 \times 2$ MIMO interference
network, and making simplifying assumptions on the channel models
if needed, future work could include the discovery of new encoding and decoding
principles inspired by the goal of characterizing information theoretic
metrics that are finer than the degrees of freedom, such as, for instance,
the generalized degrees of freedom, as was done for the two-user MIMO interference channel in \cite{Karmakar2012}. There is also the potential for
the discovery of hitherto unknown encoding schemes tailored for various
models of channel uncertainty, as has been done for the MIMO interference
and the MIMO $X$ channels in \cite{Vaze2012,Ghasemi2011} under delayed
CSIT. 



\appendices{}

\section{\label{sec:Equivalence-of-DD}Equivalence of \textmd{\normalsize{}$\mathbb{D}$
and $\mathbb{D}_{\mathrm{eq}}$}}
\begin{IEEEproof}
To make the expressions more concise, we define
\begin{gather*}
d_{{\rm sum},1}=d_{1}+d_{2}+d_{0}+d_{01}+d_{11}+d_{12}\\
d_{{\rm sum},2}=d_{1}+d_{2}+d_{0}+d_{02}+d_{21}+d_{22}\text{.}
\end{gather*}
Let $\boldsymbol{\overrightarrow{d}}=(d_{11},d_{21},d_{12},d_{22},d_{1},d_{2},d_{01},d_{02},d_{0})$.
First, prove if $\boldsymbol{\overrightarrow{d}}\in\mathbb{D}_{\mathrm{eq}}$,
then $\boldsymbol{\overrightarrow{d}}\in$$\mathbb{D}$.

Since $\boldsymbol{\overrightarrow{d}}\in\mathbb{D}_{\mathrm{eq}}$,
there exists at least a tuple $(Z_{11}$, $Z_{12}$, $Z_{21}$, $Z_{22}$,
$A_{1}$, $A_{2}$, $Z_{01}$, $Z_{02})\in\mathbb{R}_{+}^{\mathsf{A}}$
which satisfies the conditions in (\ref{eq:aux_1})-(\ref{eq:aux_12}),
such that inequalities (\ref{eq:achievable1})-(\ref{eq:achievable9})
are all satisfied. Then, from inequalities (\ref{eq:achievable1}),
(\ref{eq:aux_2}), (\ref{eq:aux_5}) and (\ref{eq:aux_6}), we get
\begin{alignat*}{1}
d_{{\rm sum},1}+d_{21}+d_{22}+d_{02} & \leq N_{1}+(M_{1}-N_{1})^{+}+d_{22}+d_{02}.
\end{alignat*}
Hence,
\begin{align*}
d_{{\rm sum},1}+d_{21}\leq N_{1}+(M_{1}-N_{1})^{+}=\max(M_{1},N_{1}),
\end{align*}
which is inequality (\ref{eq:out1}) in the definition of $\mathbb{D}$.
Similarly, it can be shown that
\begin{gather*}
d_{{\rm sum},1}+d_{22}\leq N_{1}+(M_{2}-N_{1})^{+}=\max(M_{2},N_{1})\\
d_{{\rm sum},2}+d_{11}\leq N_{2}+(M_{1}-N_{2})^{+}=\max(M_{1},N_{2})\\
d_{{\rm sum},2}+d_{12}\leq N_{2}+(M_{2}-N_{2})^{+}=\max(M_{2},N_{2}).
\end{gather*}
which are inequalities (\ref{eq:out2})-(\ref{eq:out4}) in the definition
of $\mathbb{D}$. 

Again, from inequalities (\ref{eq:achievable1}), (\ref{eq:aux_4}),
(\ref{eq:aux_5}) and (\ref{eq:aux_6}), we get
\begin{alignat*}{1}
d_{{\rm sum},1}+d_{21}+d_{22}+d_{02} & \leq N_{1}+d_{21}+d_{22}+d_{02}-A_{1},
\end{alignat*}
hence,
\begin{align*}
d_{{\rm sum},1}\leq N_{1}-A_{1}\leq N_{1},
\end{align*}
which is inequality (\ref{eq:out5}) in the definition of $\mathbb{D}$.
Similarly, we have
\begin{gather*}
d_{{\rm sum},2}\leq N_{2}-A_{2}\leq N_{2},
\end{gather*}
which is inequality (\ref{eq:out6}) in the definition of $\mathbb{D}$. 

Furthermore, inequalities (\ref{eq:out7})-(\ref{eq:out9}) hold for
$\boldsymbol{\overrightarrow{d}}$ since they are also contained in
the definition of $\mathbb{\mathbb{D}_{\mathrm{eq}}}$. Consequently,
all inequalities in the definition of $\mathbb{D}$ are satisfied
and we have that $\boldsymbol{\overrightarrow{d}}$ also belongs to
$\mathbb{D}$. Thus,
\begin{alignat}{1}
\mathbb{\mathbb{D}_{\mathrm{eq}}} & \subseteq\mathbb{D}.\label{eq:eq_in_D}
\end{alignat}

Next, we prove that if $\boldsymbol{\overrightarrow{d}}\in\mathbb{D}$,
then $\boldsymbol{\overrightarrow{d}}\in\mathbb{D}_{\mathrm{eq}}$.

For each $\boldsymbol{\overrightarrow{d}}\in\mathbb{D}$, we choose
the value for $(Z_{11}$, $Z_{12}$, $Z_{21}$, $Z_{22}$, $A_{1}$,
$A_{2}$, $Z_{01}$, $Z_{02}$) according to equations (\ref{eq:ddd1})-(\ref{eq:z_a_d_3}).
It is straightforward to verify the above choices satisfy the constraints
(\ref{eq:aux_1})-(\ref{eq:aux_12}). Also, by exhaustively enumerating
all possible relations among $M_{1},M_{2},N_{1},N_{2},d$ and removing
the $(\cdot)^{+}$ and $\min(\cdot,\cdot)$ operators, and substituting
the values of the 8 auxiliary variables, we can verify that if inequalities
(\ref{eq:out1})-(\ref{eq:out6}) hold, then inequalities (\ref{eq:achievable1})
and (\ref{eq:achievable2}) also hold. Inequalities (\ref{eq:achievable7})-(\ref{eq:achievable9})
automatically hold since they are contained in the definition of $\mathbb{D}$.
and hence $\boldsymbol{\overrightarrow{d}}$ also belongs to $\mathbb{D}_{\mathrm{eq}}$.
Thus
\begin{alignat}{1}
\mathbb{D} & \subseteq\mathbb{D}_{\mathrm{eq}}.\label{eq:d_in_eq}
\end{alignat}
Together with (\ref{eq:eq_in_D}), we have $\mathbb{D}=\mathbb{D}_{\mathrm{eq}}$.
\end{IEEEproof}
\bibliographystyle{unsrt}
\bibliography{2_user_MIMO_IC_General_Messages}

\end{document}